\documentclass[american,12pt]{article}
\usepackage{pdfpages}
\usepackage{setspace}
\usepackage{tikz}
\usepackage{epsfig}
\usepackage{amsmath}
\usepackage{amsfonts}
\usepackage{pst-plot}
\usepackage{hyperref}
\usepackage{amssymb}
\usepackage{pgffor}
\usepackage[left=1.22in, right=1.22in, top=1.25in, bottom=1.25in, includefoot, headheight=1.2in]{geometry}
\usepackage{babel}
\usepackage{subfig}
\usepackage{tikz}
\usepackage{booktabs,siunitx}
\usepackage{accents}
\usepackage{mathrsfs}
\usepackage{amsfonts}
\usepackage{amssymb}
\usepackage{pgffor}
\usepackage{eufrak}
\usepackage[utf8]{inputenc}
\usepackage{enumitem}
\usepackage{booktabs}
\usepackage{diagbox}
\usepackage[authoryear]{natbib}
\usetikzlibrary{3d}
\usepackage{pgfplots}
\usepackage{enumitem}
\usetikzlibrary{positioning}
\usetikzlibrary{decorations.pathmorphing}
\usetikzlibrary{intersections, pgfplots.fillbetween}
\usetikzlibrary{matrix}
\usepackage{tikz,tikz-3dplot}
\tdplotsetmaincoords{60}{120}
\usepackage{qtree}
\usepackage{tikz-qtree}
\usetikzlibrary{trees}
\usepackage{scalerel}
\usepackage[12pt]{moresize}
\usepackage{amsmath,fouriernc}
\usepackage{comment}

\usetikzlibrary{fpu}

\usepackage{graphicx,kantlipsum,setspace}
\usepackage{caption}
\usepackage[labelfont=bf]{caption}

\definecolor{ukrainianBlue}{rgb}{0,0.322222, 0.677778}

\definecolor{ukrainianYellow}{rgb}{255,230,0}

\newcommand*{\field}[1]{\mathbb{#1}}

\usetikzlibrary {positioning}

\definecolor {processblue}{cmyk}{0.96,0,0,0}
\usetikzlibrary{arrows}
\usepackage[authoryear]{natbib}
\input{glyphtounicode}
\DeclareMathAlphabet{\mathpzc}{OT1}{pzc}{m}{it}

\makeatletter
\renewcommand{\section}{\@startsection{section}{1}{0mm}{-1.5\baselineskip}{0.8\baselineskip}{\normalfont\large\centering}}
\renewcommand{\subsection}{\@startsection{subsection}{2}{0mm}{-0.1\baselineskip}{0.5\baselineskip}{\normalfont\bf\flushleft}}
\renewcommand{\@seccntformat}[1]{\csname the#1\endcsname \hspace{+0mm}\large{.}\hspace{+1.9mm}}
\renewcommand{\@seccntformat}[2]{\csname the#1\endcsname \hspace{+0mm}\large{.}\hspace{+1.9mm}}
\makeatother

\newtheorem{theorem}{Theorem}

\newtheorem{corollary}{Corollary}

\newtheorem{definition}{Definition}
\newtheorem{example}{Example}

\newtheorem{lemma}{Lemma}

\newenvironment{proof}[1][Proof]{\textbf{#1.} }{\ \rule{0.5em}{0.5em}}

\renewcommand{\theequation}{\arabic{equation}}

\newlength{\extraspace}
\newlength{\extraspaces}
\newcounter{dummy}

\newcommand{\baa}{
\addtocounter{equation}{1} \setcounter{dummy}{\value{equation}}
\setcounter{equation}{0}
\renewcommand{\theequation}{\arabic{dummy}\alph{equation}}
\begin{eqnarray}
\addtolength{\abovedisplayskip}{\extraspaces}
\addtolength{\belowdisplayskip}{\extraspaces}
\addtolength{\abovedisplayshortskip}{\extraspace}
\addtolength{\belowdisplayshortskip}{\extraspace}}

\newcommand{\eaa}{
\end{eqnarray}
\setcounter{equation}{\value{dummy}}
\renewcommand{\theequation}{\arabic{equation}}}

\newcommand{\be}{\begin{equation}
\addtolength{\abovedisplayskip}{\extraspaces}
\addtolength{\belowdisplayskip}{\extraspaces}
\addtolength{\abovedisplayshortskip}{\extraspace}
\addtolength{\belowdisplayshortskip}{\extraspace}}
\newcommand{\ee}{\end{equation}}

\newcommand{\ba}{\begin{eqnarray}
\addtolength{\abovedisplayskip}{\extraspaces}
\addtolength{\belowdisplayskip}{\extraspaces}
\addtolength{\abovedisplayshortskip}{\extraspace}
\addtolength{\belowdisplayshortskip}{\extraspace}}
\newcommand{\ea}{\end{eqnarray}}

\newcommand{\bd}{\begin{displaymath}
\addtolength{\abovedisplayskip}{\extraspaces}
\addtolength{\belowdisplayskip}{\extraspaces}
\addtolength{\abovedisplayshortskip}{\extraspace}
\addtolength{\belowdisplayshortskip}{\extraspace}}
\newcommand{\ed}{\end{displaymath}}

\newcommand{\deel}[2]{{\textstyle{#1 \over #2}}}
\newcommand{\hf}{{\textstyle{1\over 2}}}

\def\inbar{\,\vrule height1.5ex width.4pt depth0pt}
\font\rms=cmr12 at 12pt
\def\ce{\relax\ifmmode\mathchoice
{\hbox{$\inbar\kern-.3em{\rm C}$}} {\hbox{$\inbar\kern-.3em{\rm
C}$}} {\lower.9pt\hbox{\rms $\inbar\kern-.3em{\rm C}$}}
{\lower1.2pt\hbox{\rms $\inbar\kern-.3em{\rm C}$}}
\else{$\inbar\kern-.3em{\rm C}$}\fi}
\font\cmss=cmss12 \font\cmsss=cmss12 at 12pt
\def\ze{\relax\ifmmode\mathchoice
{\hbox{\cmss Z\kern-.4em Z}}{\hbox{\cmss Z\kern-.4em Z}}
{\lower.9pt\hbox{\cmsss Z\kern-.4em Z}} {\lower1.2pt\hbox{\cmsss
Z\kern-.4em Z}}\else{\cmss Z\kern-.4em Z}\fi}

\newcommand{\refsection}[1]{
\vspace{1mm} \pagebreak[3] \addtocounter{section}{1}
\begin{center}
{\large #1}
\end{center}
\nopagebreak
\medskip
\nopagebreak}

\def\thebibliography#1{\refsection{\bf References}
\vspace*{-8mm}\list
 {\relax}{\itemsep=1pt \parsep=0pt
 \usecounter{enumiv}\leftmargin=3em\itemindent=-\leftmargin}%
 \def\newblock{\hskip .11em plus .33em minus .07em}
 \sloppy\clubpenalty4000\widowpenalty4000
 \sfcode`\.=1000\relax}

\newcommand{\startappendix}{
\renewcommand{\thesection}{\Alph{section}}
\renewcommand{\thelemma}{\Alph{section}.\thelemma}
\setcounter{section}{0}
\setcounter{lemma}{0}
\renewcommand{\theequation}{\thesection.\arabic{equation}}}

\begin{document}

\setcounter{page}{0}
\thispagestyle{empty}

\begin{center}
{\Huge\sc Yquilibrium}\\[3mm]
{\LARGE\sc  A Theory for (Non-) Convex Economies}\\[12mm]
{\Large Jacob K. Goeree}\footnote{AGORA Center for Market Design, UNSW. I gratefully acknowledge funding from the Australian Research Council (DP190103888 and DP220102893).
I thank Brett Williams for useful suggestions.
}\\[13mm]
\today\\[10mm]
{\bf Abstract}
\end{center}
\vspace*{-3mm}
\addtolength{\baselineskip}{-1.2mm}

\begin{center}
\parbox{\textwidth}{\addtolength{\baselineskip}{-1.2mm}
\noindent General equilibrium, the cornerstone of modern economics and finance, rests on assumptions many markets do not meet. Spectrum auctions, electricity markets, and cap-and-trade programs for resource rights often feature non-convexities in preferences or production that can cause non-existence of Walrasian equilibrium and \mbox{render} general equilibrium vacuous. \mbox{\textit{Yquilibrium}} complements general equilibrium with an optimization approach to \mbox{(non-)} convex economies that does not require perfect competition. \mbox{Yquilibrium} coincides with Walrasian equilibrium when the latter \mbox{exists}. Yquilibrium exists even if Walrasian equilibrium ceases to and produces optimal \mbox{allocations} subject to linear,
anonymous, and (approximately) utility-clearing prices.}
\end{center}

\vfill

\addtocounter{footnote}{-1}

\newpage

\addtolength{\baselineskip}{1mm}

\section{Introduction}
\label{sec:intro}

A landmark of economic theory concerns the determination of prices for all goods in the economy and the demonstration that resources are efficiently allocated. General equilibrium originated with \mbox{Walras} ``\mbox{$\ldots$ whose} \mbox{system} of equations, defining equilibria in a system of interdependent quantities, is the Magna Carta of economic theory'' \citep{Schumpeter1954}.  The modern approach to general equilibrium is due to \cite{ArrowDebrue1954} and \cite{McKenzie1954} who established existence of Walrasian equilibria under convexity assumptions. \cite{DuffieSonnenschein1989} describe the history and \mbox{importance} of this existence proof, which allowed general equilibrium to gain the central role it now occupies in economics and finance.

Despite its prominence and ubiquitous use, the assumptions underlying general equilibrium limit its realism and applicability. \textit{Perfect competition} subsumes everyone optimizes taking prices as given, which begs the question ``who sets them?''\footnote{\cite{Kirman2021} provides a lucid account of the skepticism among leading twentieth century economists, e.g. Edgeworth, Hahn, Hayek, Hicks, and Pareto, about Walras' assumption of perfect competition. Walras envisioned a centralized mechanism where an auctioneer arrives at market-clearing prices via an adjustment process known as t\^atonnement. However, \cite{Scarf1960} showed that even in a simple economy with a unique equilibrium, t\^atonnement prices can cycle forever. Besides, t\^atonnement does not match the way in which prices are adjusted in the real world. Lack of convergence and realism caused interest in t\^atonnement to fade and \cite{DuffieSonnenschein1989} conclude that ``$\ldots$ few would argue its usefulness today.''} Moreover, assuming everyone submits limit orders at a common price is too restrictive. Markets typically allow participants to submit limit orders at a range of prices. In addition, they can accept others' limit orders even if these involve disequilibrium prices.

Likewise, general equilibrium's \textit{convexity} requirements are not met in many important markets. Telecoms' preferences in high-stakes spectrum auctions are typically non-convex due to synergies among the licenses for sale; virtually all producers, most notably airlines and electricity plants, face avoidable fixed costs causing non-convexities in production; many resource allocation processes, such as cap-and-trade programs for resource rights, feature participants with non-convex preferences competing for few indivisible items. Non-existence of Walrasian equilibrium, whether because of non-convex preferences, non-convexities in production, or indivisibilities, demands an alternative to complement general equilibrium's incomplete toolkit.

The shortcomings of general equilibrium in the presence of non-convexities can be illustrated with a simple exchange economy. Suppose Adam and Bob have ``max'' preferences, i.e. their utilities are $u(x,y)=\max(x,y)$. Here good 1 might be a ticket to the Opera House and good 2 a ticket to the Capitol Theatre for shows on the same evening. Adam's endowment is (2,2) and Bob's endowment is (1,1).  Walrasian equilibrium prices do not exist as Adam demands at least four units of the cheaper ticket.\footnote{Adam's income is $2p_1+2p_2$ so he demands $2+2\max(p_1,p_2)/\min(p_1,p_2)$ units of the cheaper ticket.} This does not mean, however, that gains from trade won't be seized. The Pareto-optimal outcome is for both to hold three identical tickets, e.g. Bob offers to sell his opera ticket for Adam's two theatre tickets. General equilibrium precludes this outcome since at the imputed price ratio, Adam should demand six theatre tickets in exchange for his two opera tickets. Yet, the Pareto-optimal outcome is readily accommodated by any market institution that lets Adam accept Bob's offer.

One contribution of this paper is to demonstrate that general equilibrium's assumptions are not only unrealistic they are also \textit{unnecessary}.  To this end, the paper breaks with the Walrasian paradigm that puts prices front and center, and instead adopts the Paretian perspective that the economy maximizes utility for the collective.\footnote{``The members of a collectivity enjoy maximum utility in a certain position when $\ldots$ any small displacement in departing from that position necessarily has the effect of increasing the utility which certain individuals enjoy and decreasing that which others enjoy,'' \cite{Pareto1906}.} \citeauthor{Pareto1906}'s (\citeyear{Pareto1906}) point of view is captured by the \textit{\textbf{vector maximization}} program\footnote{In economics, the dominant approach to equilibrium has always been Walrasian. If anything, Pareto's contributions are seen as part of the Walrasian paradigm. Outside of economics, Pareto's approach has gained significant traction as many problems in science and engineering involve optimization of multiple objectives. \cite{Stadler1979} surveys the historical development of vector optimization prior to 1960. In economics, research in this field was initiated by \cite{Edgeworth1881} and \cite{Pareto1906} and further developed by \cite{Koopmans1951} and \cite{KuhnTucker1951} in the fifties.}
\begin{equation}\label{P}
  \max_{\rule{0pt}{5pt}x\,\in\,F}\,\,(u_1(x_1),\ldots,u_N(x_N))\tag{P}
\end{equation}
where the allocation $x$ is the concatenation of individual consumers' bundles $x_i$, $F$ is the set of feasible allocations, and the $u_i$ are consumers' utilities. To make sense of \eqref{P} some criterion is needed to decide when one vector of utilities is better than another. Pareto's proposal is that $u$ is better than $u'$ if $u_i\geq u'_i$ for all $i$ with strict inequality for at least one $i$. Under this partial ordering, \eqref{P} yields utilities that form the Pareto frontier of the utility possibility set.

Prices are absent from Pareto's proposal. I demonstrate, however, that they follow from a dual program
\begin{equation}\label{D}
  \min_{\rule{0pt}{7pt}(p,m)\,\in\,F^*}\,\,(v_1(p,m_1),\ldots,v_N(p,m_N))\tag{D}
\end{equation}
where $F^*$ is the set of feasible prices and incomes (defined below) and the $v_i(p,m_i)$ are consumers' indirect utility functions. Specifically, in economies with only quasiconcave utilities, the dual program \eqref{D} yields indirect utilities that form the same Pareto frontier as the primal problem \eqref{P}. For such convex economies, the solutions to \eqref{P} are Pareto-optimal allocations and the solutions to \eqref{D} are Walrasian equilibrium prices (and incomes equal to the cost of the allocation at equilibrium prices). Existence of these solutions does not require fixed-point arguments, but follows since the Pareto frontiers of the programs in \eqref{P} and \eqref{D} are non-empty. The duality between \eqref{P} and \eqref{D} encapsulates the welfare theorems -- Walrasian equilibrium allocations are Pareto optimal and every Pareto-optimal allocation is part of a Walrasian equilibrium.

To summarize, the duality between \eqref{P} and \eqref{D} offers an alternative approach to equilibrium that does not rely on perfect competition. There is no assumption that consumers optimize taking prices as given, nor that prices magically solve fixed-point conditions that ensure ``market clearing.'' Instead, gains from trade are seized until an efficient outcome is reached and prices rationalize this outcome in terms of utilities. In other words, prices are ``utility clearing:'' consumers' utilities from the final allocation match the utilities they expect given prices and their initial endowments.

A second contribution of this paper is to provide an optimization approach to equilibrium. In many market-design applications, non-economic constraints such as legal and political constraints or fairness and complexity considerations play a role. These non-economic constraints supplement the usual feasibility and budget constraints and may cause non-existence of Walrasian equilibrium. In contrast, they are readily incorporated into an optimization approach.

A univariate optimization program results by ``scalarizing'' the vector optimization programs \eqref{P} and \eqref{D}. For $\alpha_i>0$ let $U_\alpha(x)=\sum_{i}\alpha_iu_i(x_i)$ denote social welfare and $V_\alpha(p,m)=\sum_i\alpha_iv_i(p,m_i)$ its dual. Their difference defines the \textbf{\textit{economy's potential}}:
\begin{equation}\label{Y}
  Y_\alpha(x,p,m)\,=\,\sum\nolimits_i\,\alpha_i(u_i(x_i)-v_i(p,m_i))\tag{Y}
\end{equation}
Below I demonstrate that in convex economies, Walrasian equilibria are maximizers of the potential for \textit{any} choice of the weights $\alpha_i$. These maxima are also roots of the potential, which reflects the utility-clearing nature of prices in convex economies.

A third contribution of this paper is to provide a general framework for non-convex economies, i.e. when not all utility functions are quasiconcave. The potential does not necessarily have roots in the presence of non-convexities as Walrasian equilibria need not exist.  The potential does, however, have maxima. This follows from Bolzano's extreme value theorem, which states that any continuous function over a compact domain will attain its maximum value at least once. The potential's maxima, called \textbf{\textit{Yquilibria}}, are the natural outcomes for non-convex economies as they entail optimal allocations subject to linear and anonymous prices that are approximately utility clearing. They correspond to Walrasian equilibria if the latter exist and are maxima of the potential $Y_\alpha$ regardless, whence the terminology Yquilibrium.

\subsection{Related Literature}

Obtaining (part of) Walrasian equilibrium via optimization has precedents. \cite{Negishi1960} obtains Pareto-optimal allocations by maximizing $U_\alpha$ and \cite{ausubel2006} obtains Walrasian prices by minimizing $V_\alpha$. The environments they consider are restrictive, however, and neither program can be extended to non-convex economies.

\citeauthor{Negishi1960}'s (\citeyear{Negishi1960}) method determines Pareto-optimal allocations as functions of the weights $\alpha_i$ by maximizing social welfare $U_\alpha$. The correct weights then follow from a system of fixed-point conditions that ensure consumers' budget constraints are met.
Maximizing the potential has several advantages compared to \citeauthor{Negishi1960}'s method. First, it produces Pareto-optimal allocations \textit{and} Walrasian prices. Second, \citeauthor{Negishi1960}'s method requires concave, rather than quasiconcave, utility functions so that the utility possibility set is convex. Example 1 shows a convex economy for which the utility possibility set is not convex and \citeauthor{Negishi1960}'s method cannot be applied.\footnote{Appendix \ref{app:dualNegishi} presents an alternative to \citeauthor{Negishi1960}'s (\citeyear{Negishi1960}) method based on minimizing a dual welfare function that is a weighted sum of indirect utilities. I demonstrate this approach can be used to derive Walrasian equilibria even when \citeauthor{Negishi1960}'s program fails.} In contrast, maximizing the potential works for any utility possibility set, convex or not. Third, the allocations and prices that follow from maximizing the potential are \textit{independent} of the $\alpha_i$. As a result, there is no need to solve a system of fixed-point conditions. Finally, \citeauthor{Negishi1960}'s method does not extend to non-convex economies.

\citeauthor{ausubel2006} (\citeyear{ausubel2006}) derives Walrasian prices by minimizing a ``Lyapunov function'' equal to the dual of social welfare $V_\alpha$.\footnote{\cite{ausubel2006} defines the Lyapunov function $L(p)=\langle p|w\rangle+\sum_iv_i(p)$ where $\langle\cdot|\cdot\rangle$ denotes the inner product, $w_k$ is the total amount of good $k$, and $v_i(p)=\max_{x_i}(u_i(x_i)-\langle p|x_i\rangle)$ is the Fenchel dual of $u_i(x_i)$. Since $w_k=\sum_i\omega_{ik}$, with $\omega_{ik}$ consumer $i$'s endowment of good $k$, it is straightforward to rewrite the Lyapunov function as $L(p)=\sum_iv_i(p,\langle p|\omega_i\rangle)$ where $v_i(p,\langle p|\omega_i\rangle)=\max_{x_i}(u_i(x_i)+\langle p|\omega_i\rangle-\langle p|x_i\rangle)$. This Lyapunov function equals dual social welfare $V_\alpha(p,m)=\sum_i\alpha_iv_i(p,m_i)$ when $\alpha_i=1$ and $m_i=\langle p|\omega_i\rangle$.} Maximizing the potential has several advantages compared to \citeauthor{ausubel2006}'s approach. First, it produces Walrasian prices \textit{and} Pareto-optimal outcomes. Second, \citeauthor{ausubel2006}'s approach requires utility functions to be concave,\footnote{For indivisible goods, \cite{ausubel2006} imposes the stronger condition that they are substitutes.} as in \cite{Negishi1960}, as well as \textit{quasilinear}. The latter assumption, while common in the auction and market-design literatures, is quite restrictive in that it rules out any income effects. Without quasilinearity, existence of a Lyapunov function that ensures price convergence cannot be guaranteed. \citeauthor{Scarf1960} (\citeyear{Scarf1960}), for instance, considers a Leontief economy with income effects and finds that prices cycle forever. In contrast, potential maximization works regardless of income effects. Finally, \citeauthor{ausubel2006}'s approach does not extend to non-convex economies.

This paper belongs to a nascent literature on non-convex market design. Following earlier work on non-convexities in general equilibrium,\footnote{\cite{Aumann1966} shows existence of Walrasian equilibrium with a continuum of traders. \cite{Starr1969} introduces the concept of \emph{quasi-equilibrium} and shows that it limits to Walrasian equilibrium with sufficiently many traders. \cite{ShapleyShubik1966} introduce the $\varepsilon$-core, which is the set of allocations that cannot be blocked by any coalition at a profit of more than $\varepsilon$. \citeauthor{ShapleyShubik1966} show that for quasilinear economies the $\varepsilon$-core exists for sufficiently many traders.} recent papers propose extensions of the Walrasian approach that produce approximately efficient and individually rational outcomes in quasilinear economies. Here approximate means that efficiency losses and individual losses become small when the number of traders, $N$, gets large. For instance, in \citeauthor{MilgromWatt2022}'s (\citeyear{MilgromWatt2022}) ``markup'' mechanism, buyers pay a markup over the sellers' price to cover payments for any unallocated supply. Building on \citeauthor{Starr1969}'s (\citeyear{Starr1969}) work, \citeauthor{MilgromWatt2022} prove the ``Bound-Form First Welfare Theorem,'' which shows that welfare losses from the markup mechanism are of order $1/N$. \citeauthor{MilgromWatt2022} show this bound also applies to a ``rationing'' mechanism in which buyers and sellers face the same price but buyers' demands may be rationed.\footnote{See also \cite{NguyenVohra2022} who prove a bound for markets with indivisible goods that depends only on a measure of preference complementarity of agents.} Both their markup and rationing mechanisms produce feasible allocations, are budget balanced, approximately individually rational, and approximately efficient.

Yquilibrium differs from these recent approaches in important ways. First, unlike the market-design literature, Yquilibrium does not impose quasilinearity. Second, Yquilibrium's properties hold for any number of traders, not just in the
limit.\footnote{Having a theory that works for any number of traders has several advantages: (i) much of our intuition, and teaching, rests on examples with few traders; (ii) it allows one to establish limit results using a model that works for any $N$ rather than approximating a theory that does not work for any $N$; (iii) there are many important non-convex markets with few players, e.g. spectrum auctions.} Third, Yquilibrium breaks with the Walrasian paradigm of price taking. In the example of Adam and Bob, no markup or rationing can yield an efficient outcome if they optimize at given prices. When the price ratio differs from one, Adam and Bob are ``on the same side of the market'' resulting in no trade. When the price ratio is one, Adam and Bob exchange one ticket, which is inefficient as Adam still holds a useless ticket. Yquilibrium predicts Adam exchanges two tickets for one of Bob's even though Adam is not ``maximizing'' at the ``given'' price ratio. To summarize, in contrast to existing market-design approaches, Yquilibrium has the following desirable features:
\begin{itemize}\addtolength{\itemsep}{-2mm}
\vspace*{-2mm}
\item[--] It allows for income effects and works for any convex or non-convex economy, not just quasilinear ones.
\item[--] It employs linear and anonymous prices (no markups) that are approximately utility clearing.
\item[--] It is budget balanced and individually rational for any $N$.
\item[--] It yields optimal allocations (without rationing) subject to linear and anonymous prices for any $N$.
\end{itemize}
Like the mechanisms considered by \cite{MilgromWatt2022}, Yquilibrium coincides with Walrasian equilibrium if the latter exists. If so, Yequilibrium prices are exactly utility clearing.

\subsection{Organization}

Section 2 provides a novel duality result in vector optimization that offers an alternative interpretation of general equilibrium and paves the way for an optimization approach to equilibrium. Section 3 applies this optimization approach to non-convex economies. Section 4 concludes. Appendix A contains proofs and Appendix B details a dual formulation of \citeauthor{Negishi1960}'s (\citeyear{Negishi1960}) method.

\section{Duality in Vector Optimization}
\label{sec:P_thm}

\noindent Consider an exchange economy with $\mathcal{N}=\{1,\ldots,N\}$ consumers and $\mathcal{K}=\{1,\ldots,K\}$ goods. For $k\in\mathcal{K}$, let $w_k>0$ denote the total amount of good $k$.
The set of feasible allocations is
\begin{displaymath}
  F(w)\,=\,\{x\in\field{R}_{\geq0}^{NK}\,|\,\sum_{i\,\in\,\mathcal{N}}x_{ik}\,\leq\,w_k\,\,\,\forall\,k\in\mathcal{K}\}
\end{displaymath}
For vectors $v,v'\in\field{R}^K$ let $\langle v|v'\rangle=\sum_{k\in\mathcal{K}}v_kv'_k$ denote the usual inner product. Without loss of generality, I normalize prices such that the economy's total income is one. The set of possible prices is then an asymmetric simplex: $\Sigma_K(w)=\{p\in\field{R}^K_{\geq 0}\,|\,\langle p|w\rangle=1\}$.

I assume that, for $i\in\mathcal{N}$, consumer $i$'s utility function $u_i:\field{R}_{\geq0}^K\rightarrow\field{R}$ is strictly increasing, quasiconcave, and differentiable. Consumer $i$'s indirect utility function
\begin{displaymath}
  v_i(p,m_i)\,=\,\max_{\rule{0pt}{6pt}\langle p|x_i\rangle\,=\,m_i}\,u_i(x_i)
\end{displaymath}
has the following properties.
\begin{lemma}\label{diewert}
The indirect utility $v_i(p,m_i)$ is
\begin{itemize}\addtolength{\itemsep}{-2mm}
\vspace*{-2mm}
\item[(i)] homogeneous of degree zero in income and prices,
\item[(ii)] non-increasing in prices and strictly increasing in income,
\item[(iii)] continuous and strictly quasiconcave (but not necessarily differentiable).
\item[(iv)] The dual of $v_i(p,m_i)$ is the utility function, i.e. $u_i(x_i)=\min_{\langle p|x_i\rangle\,=\,m_i}\,v_i(p,m_i)$.
\end{itemize}
\end{lemma}
Properties (i)-(iii) are standard, see e.g. \citeauthor{MasColell1995} (\citeyear{MasColell1995}, Prop. 3.D.3). I assume utilities are differentiable, which holds iff indirect utility is \textit{strictly} quasiconcave so that it has a unique minimizer \citep{crouzeix1983}. Property (iv) is due to \cite{diewert1974} and \cite{crouzeix1983} who use homogeneity of degree zero to normalize income to one. This is natural when studying a single consumer. However, when studying the entire economy it will prove useful to consider different income distributions.

The utility possibility set is defined as\footnote{This definition differs from the usual one that adds the negative orthant to every element of the UPS, see e.g. \citeauthor{MasColell1995} (\citeyear{MasColell1995}, p. 818). The two definitions coincide when $u_i(0)=-\infty$ for $i\in\mathcal{N}$, e.g. when $u_i(x_i)=\sum_{k\in\mathcal{K}}a_{ik}\log(x_{ik})$ for some non-negative $a_{ik}$.}
\begin{displaymath}
  \text{UPS}\,=\,\bigl\{(u_1(x_1),\ldots,u_N(x_N))\,|\,x\,\in\,F(w)\,\bigr\}
\end{displaymath}
For $u,u'\in\field{R}^N$ let $u'\geq u$ mean that $u'_i\geq u_i$ for $i\in\mathcal{N}$.  A \textbf{\textit{maximal element}} of the UPS is a vector of utilities $u\in\text{UPS}$ such that $u'\geq u$ for some $u'\in\text{UPS}$ implies $u'=u$.

I also define the \textit{indirect utility possibility set}
\begin{displaymath}
  \text{VPS}\,=\,\bigl\{(v_1(p,m_1),\ldots,v_N(p,m_N))\,|\,(p,m)\,\in\,F^*(w)\,=\,\Sigma_K(w)\times\Sigma_N\,\bigr\}
\end{displaymath}
where the incomes $m_i$ are non-negative and sum to one, i.e. $m=(m_1,\ldots,m_N)\in\Sigma_N$.
A \textbf{\textit{minimal element}} of the VPS is a vector of indirect utilities $v\in\text{VPS}$ such that $v'\leq v$ for some $v'\in\text{VPS}$ implies $v'=v$. The maximal and minimal elements exist because both the UPS and VPS have compact sections, see e.g. \citeauthor{Jahn2011} (\citeyear{Jahn2011}, Th. 6.3.c).\footnote{If for some $u\in\text{UPS}$ the set $S_u=\{u'\in\text{UPS}\,|\,u'\geq u\}$ is non-empty then $S_u$ is a section of the UPS. Likewise, if for some $v\in\text{VPS}$ the set $S_v=\{v'\in\text{VPS}\,|\,v'\leq v\}$ is non-empty then $S_v$ is a section of the VPS.}

A \mbox{\textbf{\textit{solution}}} to the vector maximization problem $\max_{x\in F}(u_1(x_1),\ldots,u_N(x_N))$ is a feasible allocation $x\in F(w)$ such that $(u_1(x_1),\ldots,u_N(x_N))$ is a maximal element of the UPS. The problem's \textbf{\textit{value}} is the set of maximal elements of the UPS, i.e. its Pareto frontier. Defining the value for a vector optimization problem is non-standard, but simplifies the statement of duality in Theorem \ref{dualVO} below. Similar definitions apply to the vector minimization problem $\min_{(p,m)\in F^*}(v_1(p,m_1),\ldots,v_N(p,m_N))$ for which the Pareto frontier is formed by the minimal elements of the VPS. The next theorem shows that the UPS and VPS intersect only along their frontiers.
\begin{theorem}\label{dualVO}
$u$ is a maximal element of the {\em UPS} iff it is a minimal element of the {\em VPS:}
\begin{equation}\label{main}
   \max_{\rule{0pt}{7pt}x\,\in\,F(w)}\,\,(u_1(x_1),\ldots,u_N(x_N))\,\,\,\,\,\,\,\,=\,\min_{\rule{0pt}{7pt}(p,m)\,\in\,F^*(w)}(v_1(p,m_1),\ldots,v_N(p,m_N))
\end{equation}
There exists exactly one solution to \eqref{main} for each $m\in\Sigma_N$, which is the unique Walrasian equilibrium of the economy with income distribution $m=(m_1,\ldots,m_N)$.
\end{theorem}
The proof of \eqref{main} is based on two lemmas, see Appendix \ref{app:proofs}. The first lemma shows that any element of the intersection $\text{UPS}\cap\text{VPS}$ must be a maximal element of the UPS and a minimal element of the VPS. In other words, the intersection $\text{UPS}\cap\text{VPS}$ does not contain interior elements of either sets. The second lemma shows that maximal elements of the UPS belong to the VPS and minimal elements of the VPS belong to the UPS, i.e. their frontiers \textit{are} in the intersection.
\begin{example}[Fenchel's economy]
\label{ex:Fenchel}
{\em
Consider an economy with two goods and two consumers with identical utility functions
\begin{equation}\label{eq:Fenchel}
  u(x,y)\,=\,x+\sqrt{y+x^2}
\end{equation}
The blue area in the left panel of Figure \ref{fig:eBox} shows the UPS, which is not convex. As a result, the Pareto-optimal allocations cannot be obtained by maximzing a weighted sum of utilities over the UPS (see also Section \ref{sec:pot}). Moreover, the usual approach of equating marginal rates of substitution to the price ratio does not apply since Pareto-optimal allocations are not necessarily interior.

\begin{figure}[t]
\begin{center}
\begin{tikzpicture}[scale=2]
\begin{scope}

\clip (0,0) rectangle (3,3);

\draw [name path=A1, very thin, ukrainianBlue, domain=0:1, samples=101, /pgf/fpu,/pgf/fpu/output format=fixed] plot ({\x^.5}, {1+(2-\x)^.5});
\draw [name path=A2, very thin, ukrainianBlue, domain=0:0.316987, samples=101, /pgf/fpu,/pgf/fpu/output format=fixed] plot ({\x+(1+\x*\x)^.5}, {2*(1-\x)});
\draw [name path=A3, very thin, ukrainianBlue, domain=0.683013:1, samples=101, /pgf/fpu,/pgf/fpu/output format=fixed] plot ({2*\x}, {1-\x+(1+(1-\x)^2)^.5});
\draw [name path=A4, very thin, ukrainianBlue, domain=0:45, samples=101] plot ({1+2^.5*cos(\x)}, {2^.5*sin(\x)});

\draw [name path=A5, very thick, red] (1+2^.5,0) -- (3,0);
\draw [name path=A6, very thick, red] (0,1+2^.5) -- (0,3);
\draw [name path=B] (0,0);
\draw [name path=C] (3,3);

\tikzfillbetween[of=A1 and B]{ukrainianBlue, opacity=0.9};
\tikzfillbetween[of=A2 and B]{ukrainianBlue, opacity=0.9};
\tikzfillbetween[of=A3 and B]{ukrainianBlue, opacity=0.9};
\tikzfillbetween[of=A4 and B]{ukrainianBlue, opacity=0.9};

\tikzfillbetween[of=A1 and C]{ukrainianYellow, opacity=1};
\tikzfillbetween[of=A2 and C]{ukrainianYellow, opacity=1};
\tikzfillbetween[of=A3 and C]{ukrainianYellow, opacity=1};
\tikzfillbetween[of=A4 and C]{ukrainianYellow, opacity=1};
\tikzfillbetween[of=A5 and C]{ukrainianYellow, opacity=1};
\tikzfillbetween[of=A6 and C]{ukrainianYellow, opacity=1};

\draw [very thick, black, domain=0.01:0.99, samples=101, /pgf/fpu,/pgf/fpu/output format=fixed] plot ({max(1/(10*(1-\x))^.5,1/(5*\x))}, {max((9/(10*(1-\x)))^.5,9/(5*\x))});
\draw [very thick, black, domain=0.01:0.99, samples=101, /pgf/fpu,/pgf/fpu/output format=fixed] plot ({max((2/(5*(1-\x)))^.5,4/(5*\x))}, {max((3/(5*(1-\x)))^.5,6/(5*\x))});
\draw [very thick, black, domain=0.01:0.99, samples=101, /pgf/fpu,/pgf/fpu/output format=fixed] plot ({max((3/(5*(1-\x)))^.5,6/(5*\x))}, {max((2/(5*(1-\x)))^.5,4/(5*\x))});
\draw [very thick, black, domain=0.01:0.99, samples=101, /pgf/fpu,/pgf/fpu/output format=fixed] plot ({max((9/(10*(1-\x)))^.5,9/(5*\x))}, {max((1/(10*(1-\x)))^.5,1/(5*\x))});

\draw [very thick, red, domain=0:1, samples=101, /pgf/fpu,/pgf/fpu/output format=fixed] plot ({\x^.5}, {1+(2-\x)^.5});
\draw [very thick, red, domain=0:0.316987, samples=101, /pgf/fpu,/pgf/fpu/output format=fixed] plot ({\x+(1+\x*\x)^.5}, {2*(1-\x)});
\draw [very thick, red, domain=0.683013:1, samples=101, /pgf/fpu,/pgf/fpu/output format=fixed] plot ({2*\x}, {1-\x+(1+(1-\x)^2)^.5});
\draw [very thick, red, domain=0:45, samples=101] plot ({1+2^.5*cos(\x)}, {2^.5*sin(\x)});

\node[scale=0.8,black] at (1.27,2.85) {$m=\deel{1}{10}$};
\node[scale=0.8,black] at (2.6,2.85) {$m=\deel{2}{5}$};
\node[scale=0.8,black] at (2.75,1.6) {$m=\deel{3}{5}$};
\node[scale=0.8,black] at (2.75,0.35) {$m=\deel{9}{10}$};

\end{scope}

\draw[->] (-0.15,0) -- (3.15,0) node[scale=0.8,below right] {$u_1$};
\draw[->] (0,-0.15) -- (0,3.15) node[scale=0.8,above left] {$u_2$};
\foreach \x/\xtext in {3/3} \draw[shift={(\x,0)}] (0pt,.2pt) -- (0pt,-.2pt) node[scale=0.7,below] {$\xtext$};
\foreach \y/\ytext in {3/3} \draw[shift={(0,\y)}] (.2pt,0pt) -- (-.2pt,0pt) node[scale=0.7,left] {$\ytext$};

\node at (5.5,1.38) {
\begin{tikzpicture}[scale=1]
\draw (0,0) -- (6,0) -- (6,6) -- (0,6) -- (0,0);
\foreach \x/\xtext in {1/\frac{1}{6}, 2/\deel{1}{3}, 3/\hf, 4/\deel{2}{3}, 5/\deel{5}{6}, 6/1} \draw[shift={(\x,0)}] (0pt,2pt) -- (0pt,-2pt) node[scale=0.7,below] {$\xtext$};
\foreach \y/\ytext in {1/\frac{1}{6}, 2/\deel{1}{3}, 3/\hf, 4/\deel{2}{3}, 5/\deel{5}{6}, 6/1} \draw[shift={(0,\y)}] (2pt,0pt) -- (-2pt,0pt) node[scale=0.7,left] {$\ytext$};

\draw [red,very thick] (0,0) -- (0,6) -- (1.90192,6) -- (4.09808,0) -- (6,0) -- (6,6);

\draw [thick, ukrainianBlue, domain=0:0.06] plot ({6*\x}, {6*(0.541534 - 4.41534*\x)});
\draw [thick, ukrainianBlue, domain=0.06:0.122648,<-] plot ({6*\x}, {6*(0.541534 - 4.41534*\x)});

\draw [thick, ukrainianBlue, domain=0:0.125] plot ({6*\x}, {6*(1 - 4*\x)});
\draw [thick, ukrainianBlue, domain=0.125:0.25,<-] plot ({6*\x}, {6*(1 - 4*\x)});

\draw [thick, ukrainianBlue, domain=0:0.25] plot ({6*\x}, {6*(1 - 2*\x)});
\draw [thick, ukrainianBlue, domain=0.25:0.5,<-] plot ({6*\x}, {6*(1 - 2*\x)});

\draw [thick, ukrainianBlue, domain=0.138751:0.35] plot ({6*\x}, {6*(1.31867 - 2.29666*\x)});
\draw [thick, ukrainianBlue, domain=0.35:0.574166,<-] plot ({6*\x}, {6*(1.31867 - 2.29666*\x)});

\draw [thick, ukrainianBlue, domain=0.425834:0.65,->] plot ({6*\x}, {6*(1.978 - 2.29666*\x)});
\draw [thick, ukrainianBlue, domain=0.65:0.861249] plot ({6*\x}, {6*(1.978 - 2.29666*\x)});

\draw [thick, ukrainianBlue, domain=0.75:0.875,->] plot ({6*\x}, {6*(4 - 4*\x)});
\draw [thick, ukrainianBlue, domain=0.875:1] plot ({6*\x}, {6*(4 - 4*\x)});

\draw [thick, ukrainianBlue, domain=0.5:0.75,->] plot ({6*\x}, {6*(2 - 2*\x)});
\draw [thick, ukrainianBlue, domain=0.75:1] plot ({6*\x}, {6*(2 - 2*\x)});

\draw [thick, ukrainianBlue, domain=00.877352:0.94,->] plot ({6*\x}, {6*(4.87381 - 4.41534*\x)});
\draw [thick, ukrainianBlue, domain=0.94:1] plot ({6*\x}, {6*(4.87381 - 4.41534*\x)});

\end{tikzpicture}};

\end{tikzpicture}
\end{center}
\vspace*{-8mm}

\caption{In the left panel, the blue area corresponds to the UPS and the yellow area to the VPS for $u_1(x,y)=u_2(x,y)=x+(y+x^2)^{1/2}$. The sets overlap only along their frontiers indicated by the red curve. The black curves show the indirect utility pairs $(v(p,1-p,m),v(p,1-p,1-m))$ for various income levels as functions of $p$. In the right panel, the red curve shows Pareto optimal allocations in the Edgeworth box and the blue lines show endowments resulting in the same Pareto optimal allocation.}\label{fig:eBox}
\vspace*{-3mm}
\end{figure}
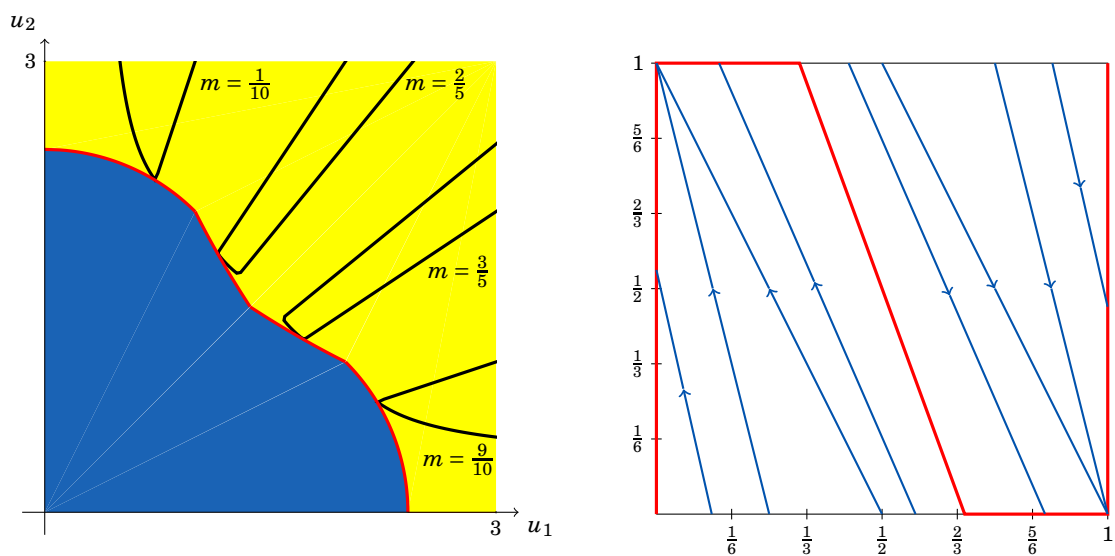

The indirect utility function is given by
\begin{displaymath}
  v(p,q,m)\,=\,\max\bigl(\frac{2m}{p},\sqrt{\frac{m}{q}}\bigr)
\end{displaymath}
The yellow area in the left panel of Figure \ref{fig:eBox} shows the VPS.
The black curves show the indirect utility pairs $(v(p,1-p,m),v(p,1-p,1-m))$ for different levels of $m$ as a function of $p\in(0,1)$. The VPS is the union of all such curves. For each $m$ there is a unique price such that the indirect utility pair belongs to the (lower) frontier of the VPS and the (upper) frontier of the UPS.

There can be no interior Pareto-optimal allocations unless $2m/p=\sqrt{m/(1-p)}$ and $2(1-m)/p=\sqrt{(1-m)/(1-p)}$, which requires $m=\hf$ and $p=\sqrt{3}-1$. This yields Pareto optimal allocations $(x_{11},x_{12})$ and $(1-x_{11},1-x_{12})$ that lie on the line
\begin{displaymath}
  x_{12}\,=\,1+\hf\sqrt{3}-(1+\sqrt{3})x_{11}
\end{displaymath}
for $\deel{1}{4}(3-\sqrt{3})\leq x_{11}\leq\deel{1}{4}(1+\sqrt{3})$.  Other Pareto-optimal allocations correspond to one consumer being indifferent while the other receives a positive amount of only one the goods.\footnote{For instance, for $0\leq m\leq\deel{1}{5}$, consumer 1 receives nothing of good 1 and $m/(1-p)$ of good 2 where the price $p$ is such that consumer 2 is indifferent, i.e. $p=2\sqrt{(2-m)(1-m)}-2(1-m)$.} The red polyline in the right panel of Figure \ref{fig:eBox} shows the Pareto-optimal allocations in the Edgeworth box. The blue lines are the budget lines for the equilibrium price and indicate endowments resulting in the same Pareto-optimal allocation.

The Walrasian equilibrium is unique for a given income distribution. The same Pareto optimal allocation can occur for different incomes, e.g. $x_1=(0,1)$ and $x_2=(1,0)$ when $\deel{1}{5}\leq m\leq\deel{1}{3}$, but the equilibrium prices differ, see the right panel of Figure \ref{fig:eBox}.
$\hfill\blacksquare$}
\end{example}

\subsection{Welfare Theorems}

The solutions to the left side of \eqref{main} are Pareto optimal allocations that yield utilities in the UPS' frontier. The solutions to the right side are incomes $m_i$ and a price vector $p$ that yield utilities $v_i(p,m_i)=u_i(x_i(p,m_i))$ that belong to the same frontier. Hence, the optimal demands are feasible and $(x,p)$ with $x=(x_1(p,m_1),\ldots,x_N(p,m_N))$ is the Walrasian equilibrium for the economy with incomes $(m_1,\ldots,m_N)$.
\begin{corollary}\label{wt1}
The duality result \eqref{main} encapsulates the first and second welfare theorem: Walrasian equilibrium allocations are Pareto optimal and any Pareto \mbox{optimal} allocation is part of a Walrasian equilibrium.
\end{corollary}
The uniqueness result in Theorem \ref{dualVO} requires strict quasiconvexity of the indirect utilities so that they have a single minimizer. Strict quasiconvexity of the indirect utility holds iff the utility function is differentiable. If this assumption is relaxed then there can be multiple Walrasian equilibria. Consider, for instance, an economy with Leontief utilities: $u_i(x,y)=\min(x,y)$ for $i=1,2$. The indirect utilities are $v_i=m_i/(p+q)$, which have linear iso-indirect-utility curves. The Walrasian equilibria for this Leontief economy are $x_1=(m,m)$ and $x_2=(1-m,1-m)$ for $m\in[0,1]$ and \textit{any} prices $p$ and $q$ that sum to one.

Assuming differentiable utilities, uniqueness of Walrasian equilibrium holds for economies parameterized by income distributions. I next consider economies parameterized by initial endowments for which multiple Walrasian equilibria may exist.

\subsection{Endowments and Equilibrium Multiplicity}

I will show that economies parameterized by endowments inherit equilibrium existence from associated economies parameterized by income distributions. To this end, I consider a more general setup where incomes are replaced by non-negative functions of prices. Specifically, consider the parameterized $(K-1)$-dimensional surface
\begin{displaymath}
  S_f\,=\,\bigl\{(v_1(p,f_1(p)),\ldots,v_N(p,f_N(p)))\,|\,p\,\in\,\Sigma_K(w)\bigr\}
\end{displaymath}
where $f=(f_1,\ldots,f_N)$ is some continuous function $f:\Sigma_K(w)\rightarrow\Sigma_N$, i.e. $f$ maps price vectors to income distributions. For instance, in Figure \ref{fig:eBox} the black curves are based on the constant functions $f_1(p)=m=1-f_2(p)$ for some constant $m$ between 0 and 1. By construction, $S_f\subset\text{VPS}$. The question is whether it intersects the UPS.

It is well known that the Walrasian equilibrium correspondence is upper-hemi-continuous, see e.g. \cite{HildenbrandMertens1972}. By Theorem \ref{dualVO} this correspondence is \mbox{single} valued when parameterized by income distributions. Hence, the map $P:\Sigma_N\rightarrow\Sigma_K(w)$, which assigns Walrasian equilibrium prices to income distributions, is both upper-hemi-continuous and single valued, i.e. it is a continuous function. By Brouwer's fixed-point theorem, $f\circ P:\Sigma_N\rightarrow\Sigma_N$ has a fixed-point.

Let $m$ denote a fixed-point of $f\circ P$ and let $p=P(m)$ then
\begin{displaymath}
  S_f\,\ni\,\bigl(v_1(p,f_1(p)),\ldots,v_N(p,f_N(p))\bigr)\,\in\,\text{UPS}\cap\text{VPS}
\end{displaymath}
i.e. $S_f$ intersects the UPS. Since $v_i(p,f_i(p))=u_i(x_i(p,m_i))$ for $i\in\mathcal{N}$, the allocation-price pair $(x,p)$ underlying this intersection is given by $x=(x_1(p,m_1),\ldots,x_N(p,m_N))$ and $p=P(m)$ with $m=(m_1,\ldots,m_N)$ a fixed-point of $f\circ P$. This allocation-price pair is the unique Walrasian equilibrium for the economy with income distribution $m$. Of course, there is no reason to assume that $f\circ P$ has a unique fixed point. In general, there can be more than one income distribution that satisfies $f(P(m))=m$, resulting in multiple Walrasian equilibria.

An example of this construction is $f_i(p)=\langle p|\omega_{i}\rangle$ for $i\in\mathcal{N}$ where $\omega_{i}\in\field{R}^K_{>0}$ denotes consumer $i$'s endowment with  $\sum_{i\in\mathcal{N}}\omega_i=w$. This choice satisfies $f_i(p)\geq 0$ and $\sum_{i\in\mathcal{N}}f_i(p)=\langle p|w\rangle=1$.
The argument of the preceding paragraph implies Walrasian equilibria exist for any choice of endowments and that they belong to $\text{UPS}\cap\text{VPS}$.
\mbox{Different} endowments can result in the same Walrasian equilibrium, see the right panel of Figure \ref{fig:eBox}, and multiple equilibria can exist. Since elements of the intersection $\text{UPS}\cap\text{VPS}$ are uniquely characterized by some $m=(m_1,\ldots,m_N)\in\Sigma_N$, each of them can be recovered by choosing $\omega_i=m_iw$ so that $f_i(p)=m_i$ for $i\in\mathcal{N}$.
\begin{corollary}\label{wt2}
For an economy parameterized by endowments there exists one or more Walrasian equilibria that belong to $\text{\em UPS}\cap\text{\em VPS}$.  Hence, any Walrasian equilibrium allocation is Pareto optimal (first welfare theorem). All Pareto optimal allocations can be obtained as part of a Walrasian equilibrium by choosing ``diagonal'' endowments $\omega_i=m_iw$ for $i\in\mathcal{N}$ and varying $m=(m_1,\ldots,m_N)$ over $\Sigma_N$ (second welfare theorem).
\end{corollary}
A Walrasian equilibrium $(x(m),p(m))$ is uniquely labeled by the income distribution, $m$. Given any initial endowment, $\omega$, a desired Walrasian equilibrium $(x(m),p(m))$ can be implemented via budget balanced income transfers $t_i=\langle p(m)|x_i(m)-\omega_i\rangle$ for $i\in\mathcal{N}$. Alternatively, consumers receive additional endowments $m_iw-\omega_i$ for $i\in\mathcal{N}$.

\subsection{The Economy's Potential}
\label{sec:pot}

One approach to vector optimization problems is to ``scalarize'' the vector objective, e.g. by taking a weighted sum of its components. \citeauthor{Negishi1960}'s (\citeyear{Negishi1960}) method, for instance,
entails maximizing utilitarian social welfare $U_\alpha(x)=\sum_{i\in\mathcal{N}}\alpha_iu_i(x_i)$ to determine
Pareto-optimal allocations as functions of the welfare weights $\alpha_i$ (assuming a convex UPS). The weights are then determined by the budget constraints. \cite{Negishi1960} shows that the correct weights are equal to the inverse of the marginal utilities of income, i.e. $\alpha_i=1/(\partial v_i/\partial m_i)$ for $i\in\mathcal{N}$.\footnote{Intuitively, these weights are such that the impact on welfare of an extra dollar to the economy is independent of who receives it.} These conditions define a system of fixed-point equations: the inverse marginal utilities of income on the right depend on prices, which in turn depend on the $\alpha_i$ weights on the left.

Unfortunately, the \citeauthor{Negishi1960} program fails if the UPS is not convex (convexifiable). This occurs when the quasiconcave utility functions cannot be ``concavified.''  A well-known example, due to Fenchel, is the utility function in \eqref{eq:Fenchel}. This utility has indifference curves that are non-parallel lines and cannot be concavified, i.e. there is no monotonic transformation of the utility function such that it becomes concave. As a result, \citeauthor{Negishi1960}'s (\citeyear{Negishi1960}) method cannot be applied. Appendix \ref{app:dualNegishi} shows that minimizing dual welfare  $V_\alpha(p,\omega)=\sum_{i\in\mathcal{N}}\alpha_iv_i(p,\langle p|\omega_i\rangle)$ with respect to prices continues to work even when \citeauthor{Negishi1960}'s program fails.

Here I follow a different route that determines prices and allocations simultaneously without having to solve for fixed-points. The \textbf{\textit{economy's potential}} is the difference between utilitarian welfare $U_\alpha(x)$ and its dual $V_\alpha(p,\omega)$:
\begin{equation}\label{Ypot}
  Y_\alpha(x,p,\omega)\,=\,\sum_{i\,\in\,\mathcal{N}}\,\alpha_i\bigl(u_i(x_i)-v_i(p,\langle p|\omega_i\rangle)\bigr)
\end{equation}
where $\alpha_i>0$ for $i\in\mathcal{N}$. The program
\begin{equation}\label{maxProgram}
    \max_{{\rule{0pt}{7pt}x\,\in\,F(w),\,p\,\in\,\Sigma_K(w)}\atop{\rule{0pt}{7pt}\langle p|x_i\rangle\,=\,\langle p|\omega_i\rangle}}\,Y_\alpha(x,p,\omega)
\end{equation}
can be used to scalarize the duality result of Theorem \ref{dualVO}.
\begin{theorem}\label{th:root}
For any choice of the welfare weights the Walrasian equilibria are the maximizers, and roots, of the economy's potential.
\end{theorem}
The program \eqref{maxProgram} offers a convenient tool to compute Walrasian equilibrium:
\begin{itemize}\addtolength{\itemsep}{-2mm}
\setlength{\itemindent}{-4mm}
\vspace*{-2mm}
\item[--] It requires no concavification of the utility functions and works for any UPS.
\item[--] It determines allocations \textit{and} prices that are independent of the $\alpha_i$ in \eqref{Ypot}.
\item[--] It dispenses with the need for solving a system of fixed-point equations.
\end{itemize}
Since the endowments $\omega_i$ enter the objective and constraints in \eqref{maxProgram} only in the form of incomes $\langle p|\omega_i\rangle$, the solution to \eqref{maxProgram} is the same for all endowments that yield the same incomes at the equilibrium price. For convex economies these endowments lie on a plane, see e.g. the blue lines in the right panel of Figure \ref{fig:eBox}. In non-convex economies this is not necessarily the case because of binding individual rationality constraints. 

Importantly, \eqref{maxProgram} offers an interpretation of the economic system different from the Walrasian paradigm. \textit{There is no assumption that consumers optimize at given prices.} Instead, \eqref{maxProgram} operationalizes the Paretian view that gains from trade are seized until an efficient allocation is reached and prices rationalize the final allocation terms utilities. Prices are ``utility clearing:'' consumers' (direct) utilities from the final allocation match the (indirect) utilities they expect given prices and their initial endowments. This alternative interpretation facilitates a generalization of \eqref{maxProgram} to non-convex economies for which Walrasian equilibria need not exist.

\section{Non-Convex Market Design}

To apply \eqref{maxProgram} to non-convex economies requires some adaptation. The first change concerns allocations. For convex economies there is no loss of generality if in \eqref{maxProgram} the allocations are restricted to the Edgeworth contract surface $E(\omega)$.\footnote{Recall that the Edgeworth contract surface consists of allocations that are Pareto optimal and individually rational, i.e. $E(\omega)=\{\,x\in F(\omega)|u(x)\geq u(\omega)\mbox{ and }u(x')\geq u(x)\mbox{ for }x'\in F(\omega)\Rightarrow x'=x\}$.} For non-convex economies this is not necessarily true as linear prices and individual rationality may limit the possible allocations. An alternative definition for the Edgeworth contract surface is needed.
The second change concerns prices. For convex economies, minimizing the dual welfare function yields utility-clearing prices. For non-convex economies the indirect utilities associated with the non-quasiconcave utilities are typically ``too large'' and produce the wrong prices. For this reason, the indirect utility functions should be associated with \textbf{\textit{quasiconcavified}} versions of the utilities. A final change concerns the weights $\alpha_i$ in \eqref{Ypot}. For convex economies, the potential's maximizers correspond to Walrasian equilibria for any choice of the weights. For non-convex economies the maxima may vary with the weights. For ease of presentation, I set $\alpha_i=1$ for $i\in\mathcal{N}$ and comment on the possibility of varying the weights in the Conclusion.

\subsection{Edgeworth's Contract Surface}
\label{subsec:Edegeworth}

The set of feasible allocations consistent with linear and anonymous prices is
\begin{displaymath}
  \overline{F}(\omega)\,=\,\{\,x\,\in\,F(\omega)\,|\,\exists\,p\,\in\,\Sigma_K(w)\,:\,\langle p|x_i\rangle\,=\,\langle p|\omega_i\rangle\,\forall\,i\,\in\,\mathcal{N}\,\}
\end{displaymath}
and
\begin{displaymath}
  \overline{E}(\omega)\,=\,\{\,x\,\in\,\overline{F}(\omega)\,|\,u(x)\,\geq\,u(\omega)\,\mbox{ and }\,u(x')\,\geq\,u(x)\,\mbox{ for }\,x\,\in\,\overline{F}(\omega)\,\Rightarrow\,x'\,=\,x\,\}
\end{displaymath}
is the set of optimal and individually rational allocations consistent with linear prices. With two consumers, $\overline{E}(\omega)=E(\omega)$, i.e. any individually rational Pareto-optimal allocation is consistent with linear and anonymous prices.
To see this, let $x=(x_1,x_2)\in E(\omega)$ with $x\neq\omega$ since otherwise any $p\in\Sigma_K$ satisfies $\langle p|x_1-\omega_1\rangle=\langle p|x_2-\omega_2\rangle=0$. For $x$ to be individually rational for both consumers, $x_1-\omega_1$ must be strictly positive for at least one good and strictly negative for at least one good. Hence, there exists $p\in\Sigma_K$ such that $\langle p|x_1-\omega_1\rangle=0$. But $x_2-\omega_2=\omega_1-x_1$ then implies $\langle p|x_2-\omega_2\rangle=0$. With more than two consumers $\overline{E}(\omega)$ need not be equal to $E(\omega)$.
\begin{example}{\em
Consider three consumers with $u_1(x,y)=\max(2x,y)$ and $\omega_1=(\deel{1}{10},\deel{1}{10})$, $u_2(x,y)=\max(x,2y)$ and $\omega_2=(\deel{4}{5},\deel{1}{10})$, and $u_3(x,y)=\min(x,y)$ and $\omega_3=(\deel{1}{10},\deel{4}{5})$. Then
\begin{displaymath}
  E(\omega)\,=\,\{\,x_1\,=\,(\alpha,0),\,x_2\,=\,(0,\alpha),\,x_3\,=\,(1-\alpha,1-\alpha)\,|\,\deel{2}{5}\,\leq\,\alpha\,\leq\,\deel{9}{10}\,\}
\end{displaymath}
No element of $E(\omega)$ is consistent with linear pricing. Consumer 1's allocation requires the price ratio $p/(1-p)$ to lie between $\deel{1}{8}$ and $\deel{1}{3}$ while consumer 2's allocation requires this ratio to lie between $\deel{3}{8}$ and 1. Individually rational optimal allocations consistent with linear and anonymous prices include
\begin{displaymath}
  \overline{E}(\omega)\,\supset\,\{\,x_1\,=\,(\beta,0),\,x_2\,=\,(0,\alpha),\,x_3\,=\,(1-\beta,1-\alpha)\,|\,\deel{2}{5}\,\leq\,\alpha\,\leq\,\deel{9}{10},\,\beta\,=\,\deel{10\alpha+7}{100\alpha-10}\,\}
\end{displaymath}
and the price ratio $p/(1-p)$ ranges from $\deel{3}{8}$ to 1. Since $\deel{1}{5}\leq\beta\leq\deel{11}{30}$, allocations in $\overline{E}(\omega)$ are not Pareto optimal as consumer 3 holds excess units of good 1 that would benefit consumer 1. However, these excess units cannot be assigned to consumer 1 in a manner consistent with linear pricing. $\hfill\blacksquare$}
\end{example}

\subsection{Quasiconcavification}

\begin{definition}
Let $u_i(x_i):F(w)\rightarrow\field{R}$ be continuous and non-decreasing and let
\begin{displaymath}
  \overline{v}_i(p,m_i)\,=\,\max_{x_i\,\in\,F(w)\,:\,\langle p|x_i\rangle\,=\,m_i}\,u_i(x_i)
\end{displaymath}
be the associated indirect utility. The quasiconcavified utility $\overline{u}_i(x_i):F(w)\rightarrow\field{R}$ is
\begin{displaymath}
  \overline{u}_i(x_i)\,=\,\min_{\langle p|x_i\rangle\,=\,m_i}\,\overline{v}_i(p,m_i)
\end{displaymath}
\end{definition}
The domain of $u_i$ is restricted to $F(w)$ to avoid that its quasiconcavification $\overline{u}_i$ is ``too large,'' see Example \ref{quasiconc}.
It is standard to show $\overline{u}_i(x_i)$ is continuous and quasiconcave for the same reason that the indirect utility is continuous and quasiconvex, e.g. \citeauthor{MasColell1995} (\citeyear{MasColell1995}, Prop. 3.D.3). The following result is due to \cite{Crouzeix1982}.
\begin{lemma}
$\overline{u}_i(x_i)$ is the smallest quasiconcave function such that $\overline{u}_i(x_i)\geq u_i(x_i)$ for all $x_i\in F(w)$. In particular, $\overline{u}_i(x_i)=u_i(x_i)$ for all $x_i\in F(w)$ iff $u_i(x_i)$ is quasiconcave. The indirect utility associated with $\overline{u}_i(x_i)$ is $\overline{v}_i(p,m)$.
\end{lemma}
While $u_i$ and $\overline{u}_i$ have the same indirect utility function, their maximizers can differ. If $x_i$ with $\langle p|x_i\rangle=m$ maximizes $u_i(x_i)$ then it maximizes $\overline{u}_i(x_i)$, but the converse is not necessarily true. Intuitively, the quasiconcavified utility $\overline{u}_i$ can have a set of maximizers of which only the extremal elements are maximizers of $u_i$.

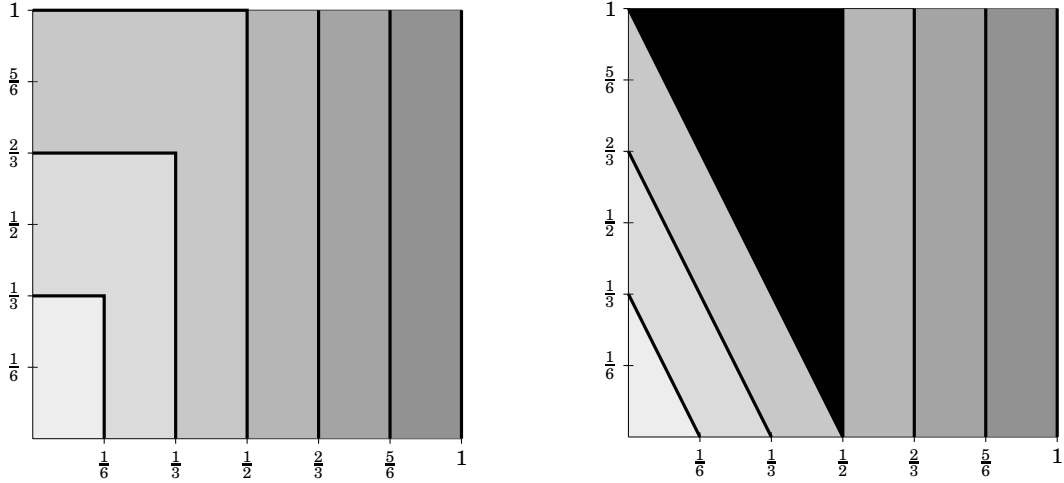
\begin{figure}[t]
\begin{center}
\begin{tikzpicture}[scale=0.95]
\draw (0,0) -- (6,0) -- (6,6) -- (0,6) -- (0,0);
\foreach \x/\xtext in {1/\frac{1}{6}, 2/\deel{1}{3}, 3/\hf, 4/\deel{2}{3}, 5/\deel{5}{6}, 6/1} \draw[shift={(\x,0)}] (0pt,2pt) -- (0pt,-2pt) node[scale=0.7,below] {$\xtext$};
\foreach \y/\ytext in {1/\frac{1}{6}, 2/\deel{1}{3}, 3/\hf, 4/\deel{2}{3}, 5/\deel{5}{6}, 6/1} \draw[shift={(0,\y)}] (2pt,0pt) -- (-2pt,0pt) node[scale=0.7,left] {$\ytext$};

\draw [name path=A] (0,0);
\draw [name path=B,gray, opacity=1/7] (1,0) -- (1,2) -- (0,2);
\draw [name path=C,gray, opacity=2/7] (2,0) -- (2,4) -- (0,4);
\draw [name path=D,gray, opacity=3/7] (3,0) -- (3,6) -- (0,6);
\draw [name path=DD,gray, opacity=3/7] (3,0) -- (3,6);
\draw [name path=E,gray, opacity=4/7] (4,0) -- (4,6);
\draw [name path=F,gray, opacity=5/7] (5,0) -- (5,6);
\draw [name path=G,gray, opacity=6/7] (6,0) -- (6,6);
\tikzfillbetween[of=A and B]{gray, opacity=1/7};
\tikzfillbetween[of=B and C]{gray, opacity=2/7};
\tikzfillbetween[of=C and D]{gray, opacity=3/7};
\tikzfillbetween[of=DD and E]{gray, opacity=4/7};
\tikzfillbetween[of=E and F]{gray, opacity=5/7};
\tikzfillbetween[of=F and G]{gray, opacity=6/7};
\draw [very thick, black, opacity=1] (1,0) -- (1,2) -- (0,2);
\draw [very thick, black, opacity=1] (2,0) -- (2,4) -- (0,4);
\draw [very thick, black, opacity=1] (3,0) -- (3,6) -- (0,6);
\draw [very thick, black, opacity=1] (4,0) -- (4,6);
\draw [very thick, black, opacity=1] (5,0) -- (5,6);
\draw [very thick, black, opacity=1] (6,0) -- (6,6);

\node at (11.2,2.8) {
\begin{tikzpicture}[scale=0.95]
\draw (0,0) -- (6,0) -- (6,6) -- (0,6) -- (0,0);
\foreach \x/\xtext in {1/\frac{1}{6}, 2/\deel{1}{3}, 3/\hf, 4/\deel{2}{3}, 5/\deel{5}{6}, 6/1} \draw[shift={(\x,0)}] (0pt,2pt) -- (0pt,-2pt) node[scale=0.7,below] {$\xtext$};
\foreach \y/\ytext in {1/\frac{1}{6}, 2/\deel{1}{3}, 3/\hf, 4/\deel{2}{3}, 5/\deel{5}{6}, 6/1} \draw[shift={(0,\y)}] (2pt,0pt) -- (-2pt,0pt) node[scale=0.7,left] {$\ytext$};

\draw [name path=A] (0,0);
\draw [name path=B,gray, opacity=1/7] (1,0) -- (0,2);
\draw [name path=C,gray, opacity=2/7] (2,0) -- (0,4);
\draw [name path=D,gray, opacity=3/7] (3,0) -- (0,6);
\draw [name path=E,gray, opacity=3/7] (3,0) -- (3,6) -- (0,6);
\draw [name path=F,gray, opacity=4/7] (4,0) -- (4,6);
\draw [name path=G,gray, opacity=5/7] (5,0) -- (5,6);
\draw [name path=H,gray, opacity=6/7] (6,0) -- (6,6);
\tikzfillbetween[of=A and B]{gray, opacity=1/7};
\tikzfillbetween[of=B and C]{gray, opacity=2/7};
\tikzfillbetween[of=C and D]{gray, opacity=3/7};
\tikzfillbetween[of=E and D]{black, opacity=1};
\tikzfillbetween[of=F and E]{gray, opacity=4/7};
\tikzfillbetween[of=G and F]{gray, opacity=5/7};
\tikzfillbetween[of=H and G]{gray, opacity=6/7};
\draw [very thick, black, opacity=1] (1,0) -- (0,2);
\draw [very thick, black, opacity=1] (2,0) -- (0,4);
\draw [very thick, black, opacity=1] (3,0) -- (0,6);
\draw [very thick, black, opacity=1] (3,0) -- (3,6);
\draw [very thick, black, opacity=1] (4,0) -- (4,6);
\draw [very thick, black, opacity=1] (5,0) -- (5,6);
\draw [very thick, black, opacity=1] (6,0) -- (6,6);

\end{tikzpicture}};

\end{tikzpicture}
\end{center}
\vspace*{-6mm}

\caption{The left panel shows the indifference curves for $u_2(x,y)=\max(2x,y)$ and the right panel shows the indifference curves for its quasiconcavification $\overline{u}_2(x,y)=\max(\min(2x+y,1),2x)$. The latter is non-decreasing and quasiconcave with a ``thick'' indifference curve at $\overline{u}_2=1$.}\label{fig:concMax}
\vspace*{0mm}
\end{figure}

\begin{example}\label{quasiconc}
{\em
Let $u(x,y)=\max(2x,y)$ with $0\leq x,y\leq 1$. The concavified utility $2x+y$ is ``too large,'' e.g. it generates a utility of 3 when $x=y=1$, while the original utility is capped at 2. Figure \ref{fig:concMax} shows the indifference map for $u(x,y)$ on the left and for its quasiconcavification
\begin{displaymath}
  \overline{u}(x,y)\,=\,\max\bigl(\min(2x+y,1),2x\bigr)
\end{displaymath}
on the right. The latter is non-decreasing and quasiconcave with a ``thick'' indifference curve at $\overline{u}=1$. The indirect utility function associated with $\overline{u}$ is
\begin{displaymath}
  \overline{v}(p,q,m)\,=\,\max\bigl(2\min\bigl(1,\frac{m}{p}\bigr),\min\bigl(1,\frac{m}{q}\bigr)\bigr)
\end{displaymath}
which is smaller than the indirect utility $\max(2m/p,m/q)$ associated with $2x+y$.
$\hfill\blacksquare$}
\end{example}

\subsection{Yquilibrium}
\label{sec:Yeq}

Incorporating the above adaptations into \eqref{maxProgram} yields a maximization approach to equilibrium that applies to convex and non-convex economies alike.
\begin{definition}\label{def:Y}
A Yquilibrium is an allocation-price pair $(x,p)$ that solves
\begin{equation}\label{pot}
    \max_{\rule{0pt}{7pt}x\,\in\,\overline{E}(\omega),\,p\,\in\,\Sigma_K(w)}\,Y(x,p,\omega)
\end{equation}
where the potential is given by $Y(x,p,\omega)=\sum_{i\in\mathcal{N}}u_i(x_i)-\overline{v}_i(p,\langle p|\omega_i\rangle)$.
\end{definition}
The next theorem shows that for convex economies there are no additional Yquilibria besides the Walrasian equilibria. Importantly, for a non-convex economy, Yquilibrium exists even when Walrasian equilibrium ceases to.
\begin{theorem}\label{potential}
If an economy has a Walrasian equilibrium then its Yquilibria coincide with its Walrasian equilibria and the value of the program in \eqref{pot} is nil. Otherwise, the value of \eqref{pot} may be negative, and the economy's Yquilibria consist of optimal allocations consistent with linear, anonymous, and (approximate) utility-clearing prices.
\end{theorem}
Recall that $\overline{E}(\omega)=E(\omega)$ when there are two consumers.
\begin{corollary}
Yquilibrium allocations are Pareto optimal in the case of bilateral trade.
\end{corollary}
These Pareto-optimal allocations are supported by linear and anonymous prices, which are not necessarily Walrasian prices if utilities are not quasiconcave.
\begin{example}
{\em
Consider an economy with two-goods, one unit of each, and two consumers with utilities $u_1(x,y)=x^{2/3}y^{1/3}$ and $u_2(x,y)=\max(2x,y)$. The left panel of Figure \ref{fig:UPS-NC} shows the UPS in blue and the VPS in yellow. Their frontiers do not overlap except in a single point where consumer 1 owns all the goods.  Hence, no Walrasian equilibrium exists when consumer 2's endowment is non-trivial.

\begin{figure}[t]
\begin{center}
\begin{tikzpicture}[scale=6]

\begin{scope}

\clip (0,0) rectangle (1,1);

\draw [name path=A, very thick, ukrainianBlue, domain=0:1, samples=101] plot ({\x}, {(2/3*(1-\x^1.5)});
\draw [name path=B] (0,0);
\tikzfillbetween[of=A and B]{ukrainianBlue, opacity=0.9};

\draw [name path=C, very thick, ukrainianYellow] (1,0) -- (0,1);
\draw [name path=D] (1,1);
\tikzfillbetween[of=C and D]{ukrainianYellow, opacity=1};

\draw [very thick, black, domain=0.33:0.8, samples=101, /pgf/fpu,/pgf/fpu/output format=fixed] plot ({1/9*2^(2/3)*1/(\x^2*(1-\x))^(1/3)}, {1/3*max(4/(3*\x),2/(3*(1-\x)))});
\draw [very thick, black, domain=0.2:0.9, samples=101, /pgf/fpu,/pgf/fpu/output format=fixed] plot ({1/6*2^(2/3)*1/(\x^2*(1-\x))^(1/3)}, {1/3*max(1/(\x),1/(2*(1-\x)))});
\draw [very thick, black, domain=0.2:0.9, samples=101, /pgf/fpu,/pgf/fpu/output format=fixed] plot ({2/9*2^(2/3)*1/(\x^2*(1-\x))^(1/3)}, {1/3*max(2/(3*\x),1/(3*(1-\x)))});

\node[scale=0.8,black] at (0.25,0.95) {$m=\deel{1}{3}$};
\node[scale=0.8,black] at (0.7,0.95) {$m=\deel{1}{2}$};
\node[scale=0.8,black] at (0.92,0.63) {$m=\deel{2}{3}$};

\end{scope}

\draw[->] (-0.05,0) -- (1.05,0) node[scale=0.8,below right] {$u_1$};
\draw[->] (0,-0.05) -- (0,1.05) node[scale=0.8,above left] {$u_2$};
\foreach \x/\xtext in {1/1} \draw[shift={(\x,0)}] (0pt,.2pt) -- (0pt,-.2pt) node[scale=0.7,below] {$\xtext$};
\foreach \y/\ytext in {0.333/1, 0.667/2, 1/3} \draw[shift={(0,\y)}] (.2pt,0pt) -- (-.2pt,0pt) node[scale=0.7,left] {$\ytext$};

\node at (1.84,0.46) {
\begin{tikzpicture}[scale=1]

\begin{scope}

\clip (0,0) rectangle (6,6);

\filldraw[white, opacity=1] (0,0) -- (6*3/4,0) -- (6*1/2,6) -- (0,6) -- (0,0);
\draw [name path=A, gray, opacity=1] (6*3/4,0) -- (6*1/2,6);
\draw [name path=B, gray, opacity=1, domain=0.5:1] plot ({6*\x}, {6*min(1/(4*\x^2),2-2*\x)});
\tikzfillbetween[of=A and B]{gray, opacity=.15};
\draw [name path=C, line width=0.01pt, gray, opacity=0.1, domain=0.5:1] plot ({6*\x}, {6*max(1/(4*\x^2),(2*\x-1)^2)});
\draw [name path=D, line width=0.01pt, red] (3,6) -- (6,6);
\tikzfillbetween[of=C and D]{gray, opacity=.45};
\draw [name path=E, line width=0.01pt, gray, opacity=0.1, domain=0.809017:1] plot ({6*\x}, {6*(2*\x-1)^2});
\draw [name path=F, line width=0.01pt, gray, opacity=0.1] (6*0.809017,6*0.381966) -- (6,0);
\tikzfillbetween[of=E and F]{gray, opacity=.3};

\draw [thick, ukrainianBlue, ->] (6*1/2,0) -- (6*11/24,6*1/4);
\draw [thick, ukrainianBlue] (6*11/24,6*1/4) -- (6*1/3,6*1);

\draw [thick, ukrainianBlue, ->] (6*1/4,0) -- (6*5.5/24,6*1/4);
\draw [thick, ukrainianBlue] (6*5.5/24,6*1/4) -- (6*1/6,6*1);

\draw [thick, ukrainianBlue, ->] (6*1/10,0) -- (6*5.5/60,6*1/4);
\draw [thick, ukrainianBlue] (6*5.5/60,6*1/4) -- (6*1/15,6*1);

\draw [thick, ukrainianBlue, ->] (6*3/4,0) -- (6*5.5/8,6*1/4);
\draw [thick, ukrainianBlue] (6*5.5/8,6*1/4) -- (6*1/2,6*1);

\draw [name path=C1, thick, ukrainianBlue,->] (6,0) -- (6*0.875,6*0.25);
\draw [name path=C2, thick, ukrainianBlue] (6*0.875,6*0.25) -- (6*0.809017,6*0.381966);
\draw [name path=C3, thick, ukrainianBlue, domain=0.5:0.809017] plot ({6*\x}, {6/(2*\x)^2});

\draw [thick, ukrainianBlue,->] (7,0) -- (6*0.974991,6*0.383352);
\draw [thick, ukrainianBlue] (6*0.974991,6*0.383352) -- (6*0.879153,6*0.575028);
\draw [thick, ukrainianBlue, domain=2/3:0.879153] plot ({6*\x}, {6*(4/(6*\x))^2});

\draw [thick, ukrainianBlue,->] (8,0) -- (6*0.98133,6*0.704006);
\draw [thick, ukrainianBlue] (6*0.98133,6*0.704006) -- (6*0.942219,6*0.782229);
\draw [thick, ukrainianBlue, domain=5/6:0.942219] plot ({6*\x}, {6*(5/(6*\x))^2});

\draw [thick, red] (0.01,0) -- (0.01,6-0.01) -- (6,6-0.01);

\node at (2,1) {$\text{I}$};
\node at (4.8,1) {$\text{II}$};
\node at (5.45,2) {$\text{III}$};
\node at (4.95,5) {$\text{IV}$};

\end{scope}

\draw (0,0) -- (6,0) -- (6,6) -- (0,6) -- (0,0);
\foreach \x/\xtext in {1/\frac{1}{6}, 2/\deel{1}{3}, 3/\hf, 4/\deel{2}{3}, 5/\deel{5}{6}, 6/1} \draw[shift={(\x,0)}] (0pt,2pt) -- (0pt,-2pt) node[scale=0.7,below] {$\xtext$};
\foreach \y/\ytext in {1/\frac{1}{6}, 2/\deel{1}{3}, 3/\hf, 4/\deel{2}{3}, 5/\deel{5}{6}, 6/1} \draw[shift={(0,\y)}] (2pt,0pt) -- (-2pt,0pt) node[scale=0.7,left] {$\ytext$};

\end{tikzpicture}};

\end{tikzpicture}
\end{center}
\vspace*{-6mm}

\caption{The blue area in the left panel shows the UPS and the yellow area the VPS for $u_1(x,y)=x^{2/3}y^{1/3}$ and $u_2(x,y)=\max(2x,y)$. The black curves show indirect utility pairs $(v_1(p,1-p,m),v_2(p,1-p,1-m))$ for various income levels as functions of $p$. In the right panel, the top edge (left edge) of the Edgeworth box shows (weakly) Pareto optimal allocations. The blue lines correspond to endowments that yield the same Pareto optimal allocation. There are four different price regimes as indicated by the Roman numerals I--IV and the different coloring.}\label{fig:UPS-NC}
\vspace*{-2mm}
\end{figure}
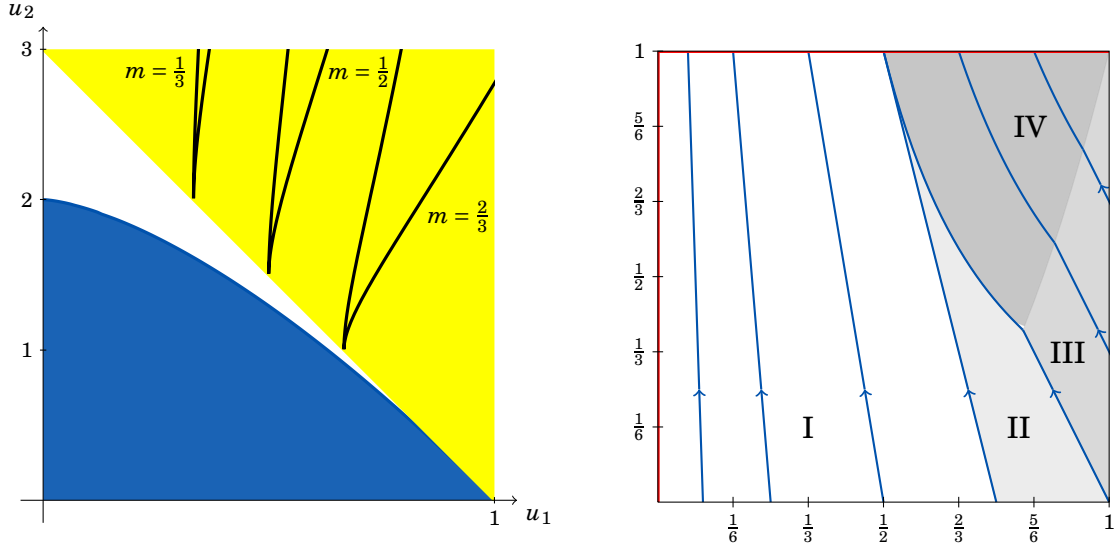

Corollary 3 implies that from any interior endowment, trade will lead to a Pareto-optimal allocation at the top of the Edgeworth box. These allocations can be parameterized as
\begin{displaymath}
  x_1(p,\omega)\,=\,\bigl(\omega_{11}-\frac{1-p}{p}(1-\omega_{12}),1\bigr)
\end{displaymath}
and $x_2(p,\omega)=(1,1)-x_1(p,\omega)$, where $(p,1-p)$ is the price vector for $0<p<1$. The program in \eqref{pot} reduces to
\begin{displaymath}
  \max_{\rule{0pt}{7pt}\frac{1+\sqrt{\omega_{12}}}{1+\omega_{11}+\sqrt{\omega_{12}}}\,\leq\,p\,\leq\,1}\,Y\bigl(x(p,\omega),p,\omega\bigr)
\end{displaymath}
where the lower bound for $p$ ensures individual rationality for consumer 1. This is a simple constrained optimization problem in one variable, $p$. The optimal price takes a different form across four regions, labeled I to IV, that result from binding individual rationality constraints.\footnote{The regions can be parameterized by consumer 1's endowments $(\omega_{11},\omega_{12})$ with $0\leq\omega_{11},\omega_{12}\leq 1$. They are $\text{I}=\{\omega_{11}\leq\deel{3}{4},\,\omega_{12}\leq 3-4\omega_{11}\}$, $\text{II}=\{\hf\leq\omega_{11}\leq 1,\,3-4\omega_{11}\leq\omega_{12}\leq\min((2\omega_{11})^{-2},2-2\omega_{11})\}$, \mbox{$\text{III}=\{\omega_{11}\geq\deel{1}{4}(1+\sqrt{5}),\,2-2\omega_{11}\leq\omega_{12}\leq(2\omega_{11}-1)^2\}$,} and
$\text{IV}=\{\omega_{11}\geq\hf,\,\omega_{12}\geq\max((2\omega_{11})^{-2},(2\omega_{11}-1)^2)\}$.} The Yquilibrium price is
\begin{displaymath}
  p\,=\,\left\{\begin{array}{lll}
  \frac{3-\omega_{12}}{3+\omega_{11}-\omega_{12}} & \text{in region I} \\[2mm]
  \frac{2-2\omega_{12}}{1+2\omega_{11}-2\omega_{12}} & \text{in region II} \\[2mm]
  \frac{2}{3} & \text{in region III} \\[2mm]
  \frac{1+\sqrt{\omega_{12}}}{1+\omega_{11}+\sqrt{\omega_{12}}} & \text{in region IV}
  \end{array}\right.
\end{displaymath}
and the corresponding Yquilibrium allocation is
\begin{displaymath}
  x_1(p,\omega)\,=\,\left\{\begin{array}{lll}
  (\frac{2\omega_{11}}{3-\omega_{12}},1) & \text{in region I} \\[2mm]
  (\frac{1}{2},1) & \text{in region II}  \\[2mm]
  (\omega_{11}+\hf\omega_{12}-\hf,1) & \text{in region III} \\[2mm]
  (\omega_{11}\sqrt{w_{1y}},1) & \text{in region IV}
  \end{array}\right.
\end{displaymath}
This price-allocation pair produces the pattern in the right panel of Figure \ref{fig:UPS-NC}.

\begin{figure}[p]
\begin{center}
\begin{tikzpicture}[scale=6]

\begin{scope}

\clip (0,0) rectangle (1,1);

\draw [name path=A, very thick, ukrainianBlue] (0,0) -- (1,0) -- (0,1) -- (0,0);
\draw [name path=B] (0,0);
\tikzfillbetween[of=A and B]{ukrainianBlue, opacity=0.9};

\draw [name path=C, very thick, ukrainianYellow] (1,0) -- (0,1);
\draw [name path=D] (1,1);
\tikzfillbetween[of=C and D]{ukrainianYellow, opacity=1};

\draw [very thick, black, domain=0.33:0.8, samples=101, /pgf/fpu,/pgf/fpu/output format=fixed] plot ({1/9*2^(2/3)*1/(\x^2*(1-\x))^(1/3)}, {1/3*max(4/(3*\x),2/(3*(1-\x)))});
\draw [very thick, black, domain=0.2:0.9, samples=101, /pgf/fpu,/pgf/fpu/output format=fixed] plot ({1/6*2^(2/3)*1/(\x^2*(1-\x))^(1/3)}, {1/3*max(1/(\x),1/(2*(1-\x)))});
\draw [very thick, black, domain=0.2:0.9, samples=101, /pgf/fpu,/pgf/fpu/output format=fixed] plot ({2/9*2^(2/3)*1/(\x^2*(1-\x))^(1/3)}, {1/3*max(2/(3*\x),1/(3*(1-\x)))});

\node[scale=0.8,black] at (0.25,0.95) {$m=\deel{1}{3}$};
\node[scale=0.8,black] at (0.7,0.95) {$m=\deel{1}{2}$};
\node[scale=0.8,black] at (0.92,0.63) {$m=\deel{2}{3}$};

\end{scope}

\draw [very thick, red] (1,0) -- (0,1);

\draw[->] (-0.05,0) -- (1.05,0) node[scale=0.8,below right] {$u_1$};
\draw[->] (0,-0.05) -- (0,1.05) node[scale=0.8,above left] {$u_2$};
\foreach \x/\xtext in {1/1} \draw[shift={(\x,0)}] (0pt,.2pt) -- (0pt,-.2pt) node[scale=0.7,below] {$\xtext$};
\foreach \y/\ytext in {0.333/1, 0.667/2, 1/3} \draw[shift={(0,\y)}] (.2pt,0pt) -- (-.2pt,0pt) node[scale=0.7,left] {$\ytext$};

\node at (1.84,0.46) {
\begin{tikzpicture}[scale=1]

\begin{scope}

\clip (0,0) rectangle (6,6);

\draw [thick, ukrainianBlue, ->] (1,0) -- (0.75,0.5);
\draw [thick, ukrainianBlue] (0.75,0.5) -- (0.5,1);
\draw [thick, ukrainianBlue, ->] (-2,6) -- (0.5,1);

\draw [thick, ukrainianBlue, ->] (2,0) -- (5/3,2/3);
\draw [thick, ukrainianBlue] (5/3,2/3) -- (1,2);
\draw [thick, ukrainianBlue, ->] (-1,6) -- (1,2);

\draw [thick, ukrainianBlue, ->] (3,0) -- (2.5,1);
\draw [thick, ukrainianBlue] (2.5,1) -- (1.5,3);
\draw [thick, ukrainianBlue, ->] (0,6) -- (1.5,3);

\draw [thick, ukrainianBlue, ->] (4,0) -- (3.25,1.5);
\draw [thick, ukrainianBlue] (3.25,1.5) -- (2.1,3.8);
\draw [thick, ukrainianBlue, ->] (1,6) -- (2.1,3.8);

\draw [thick, ukrainianBlue, ->] (5,0) -- (4,2);
\draw [thick, ukrainianBlue] (4,2) -- (2.8,4.4);
\draw [thick, ukrainianBlue, ->] (2,6) -- (2.8,4.4);

\draw [thick, ukrainianBlue,->] (6,0) -- (4.7,2.6);
\draw [thick, ukrainianBlue] (4.7,2.6) -- (3.55,4.9);
\draw [thick, ukrainianBlue, ->] (3,6) -- (3.55,4.9);

\draw [thick, ukrainianBlue,->] (7,0) -- (5.3,3.4);
\draw [thick, ukrainianBlue] (5.3,3.4) -- (4.375,5.25);
\draw [thick, ukrainianBlue, ->] (4,6) -- (4.375,5.25);

\draw [thick, ukrainianBlue,->] (8,0) -- (5.75,4.5);
\draw [thick, ukrainianBlue] (5.75,4.5) -- (5.25,5.5);
\draw [thick, ukrainianBlue, ->] (5,6) -- (5.25,5.5);

\draw [thick, red] (0,0) -- (6,6);

\end{scope}

\draw (0,0) -- (6,0) -- (6,6) -- (0,6) -- (0,0);
\foreach \x/\xtext in {1/\frac{1}{6}, 2/\deel{1}{3}, 3/\hf, 4/\deel{2}{3}, 5/\deel{5}{6}, 6/1} \draw[shift={(\x,0)}] (0pt,2pt) -- (0pt,-2pt) node[scale=0.7,below] {$\xtext$};
\foreach \y/\ytext in {1/\frac{1}{6}, 2/\deel{1}{3}, 3/\hf, 4/\deel{2}{3}, 5/\deel{5}{6}, 6/1} \draw[shift={(0,\y)}] (2pt,0pt) -- (-2pt,0pt) node[scale=0.7,left] {$\ytext$};

\end{tikzpicture}};

\end{tikzpicture}

\vspace*{5mm}

\begin{tikzpicture}[scale=6]

\begin{scope}

\clip (0,0) rectangle (1,1);

\draw [name path=A, very thick, ukrainianBlue, domain=0:1, samples=101] plot ({\x}, {3/2*min((1-\x),max(1/3,2/3*(1-\x^1.5))});
\draw [name path=B] (0,0);
\tikzfillbetween[of=A and B]{ukrainianBlue, opacity=0.9};

\draw [name path=C, very thin, ukrainianYellow, domain=0:1, samples=101] plot ({\x}, {3/2*min((1-\x),max(1/3,2/3*(1-\x^1.5))});
\draw [name path=D] (0,1) -- (1,1);
\tikzfillbetween[of=C and D]{ukrainianYellow, opacity=1};

\draw [very thick, red, domain=0:1, samples=101] plot ({\x}, {3/2*min((1-\x),max(1/3,2/3*(1-\x^1.5))});

\draw [very thick, black, domain=0.6:0.996, samples=101, /pgf/fpu,/pgf/fpu/output format=fixed] plot ({1/9*2^(2/3)*1/(\x^2*(1-\x))^(1/3)}, {1/2*max(2*min(1,2/(3*\x)),min(1,2/(3*(1-\x))))});
\draw [very thick, black, domain=0.5:0.99, samples=101, /pgf/fpu,/pgf/fpu/output format=fixed] plot ({1/6*2^(2/3)*1/(\x^2*(1-\x))^(1/3)}, {1/2*max(2*min(1,1/(2*\x)),min(1,1/(2*(1-\x))))});
\draw [very thick, black, domain=0.2:0.99, samples=101, /pgf/fpu,/pgf/fpu/output format=fixed] plot ({5/18*2^(2/3)*1/(\x^2*(1-\x))^(1/3)}, {1/2*max(2*min(1,1/(6*\x)),min(1,1/(6*(1-\x))))});

\node[scale=0.8,black] at (0.25,0.95) {$m=\deel{1}{3}$};
\node[scale=0.8,black] at (0.6,0.95) {$m=\deel{1}{2}$};
\node[scale=0.8,black] at (0.925,0.27) {$m=\deel{5}{6}$};

\end{scope}

\draw[->] (-0.05,0) -- (1.05,0) node[scale=0.8,below right] {$u_1$};
\draw[->] (0,-0.05) -- (0,1.05) node[scale=0.8,above left] {$u_2$};
\foreach \x/\xtext in {1/1} \draw[shift={(\x,0)}] (0pt,.2pt) -- (0pt,-.2pt) node[scale=0.7,below] {$\xtext$};
\foreach \y/\ytext in {0.5/1, 1/2} \draw[shift={(0,\y)}] (.2pt,0pt) -- (-.2pt,0pt) node[scale=0.7,left] {$\ytext$};

\node at (1.84,0.46) {
\begin{tikzpicture}[scale=1]

\begin{scope}

\clip (0,0) rectangle (6,6);

\draw [thick, ukrainianBlue, ->] (6*1/2,0) -- (6*11/24,6*1/4);
\draw [thick, ukrainianBlue] (6*11/24,6*1/4) -- (6*1/3,6*1);

\draw [thick, ukrainianBlue, ->] (6*1/4,0) -- (6*5.5/24,6*1/4);
\draw [thick, ukrainianBlue] (6*5.5/24,6*1/4) -- (6*1/6,6*1);

\draw [thick, ukrainianBlue, ->] (6*1/10,0) -- (6*5.5/60,6*1/4);
\draw [thick, ukrainianBlue] (6*5.5/60,6*1/4) -- (6*1/15,6*1);

\draw [thick, ukrainianBlue, ->] (6*3/4,0) -- (6*5.5/8,6*1/4);
\draw [thick, ukrainianBlue] (6*5.5/8,6*1/4) -- (6*1/2,6*1);

\draw [thick, ukrainianBlue,->] (6,0) -- (6*0.875,6*0.25);
\draw [thick, ukrainianBlue] (6*0.875,6*0.25) -- (6*0.5,6*1);

\draw [thick, ukrainianBlue,->] (7,0) -- (5.4,3.2);
\draw [thick, ukrainianBlue] (5.4,3.2) -- (4.3,5.4);
\draw [thick, ukrainianBlue,->] (4,6) -- (4.3,5.4);

\draw [thick, ukrainianBlue,->] (8,0) -- (5.6,4.8);
\draw [thick, ukrainianBlue] (5.6,4.8) -- (5.2,5.6);
\draw [thick, ukrainianBlue,->] (5,6) -- (5.2,5.6);

\draw [thick, red] (0.01,0) -- (0.01,6-0.01) -- (3,6-0.01) -- (4,4) -- (6,6);

\end{scope}

\draw (0,0) -- (6,0) -- (6,6) -- (0,6) -- (0,0);
\foreach \x/\xtext in {1/\frac{1}{6}, 2/\deel{1}{3}, 3/\hf, 4/\deel{2}{3}, 5/\deel{5}{6}, 6/1} \draw[shift={(\x,0)}] (0pt,2pt) -- (0pt,-2pt) node[scale=0.7,below] {$\xtext$};
\foreach \y/\ytext in {1/\frac{1}{6}, 2/\deel{1}{3}, 3/\hf, 4/\deel{2}{3}, 5/\deel{5}{6}, 6/1} \draw[shift={(0,\y)}] (2pt,0pt) -- (-2pt,0pt) node[scale=0.7,left] {$\ytext$};

\end{tikzpicture}};

\end{tikzpicture}
\end{center}
\vspace*{-6mm}

\caption{The blue areas in the left panels show the UPS and the yellow areas the VPS for $u_1(x,y)=x^{2/3}y^{1/3}$ and the concavified utility $\overline{u}_2(x,y)=2x+y$ (top panels) and for $u_1(x,y)=x^{2/3}y^{1/3}$ and the quasiconcavified utility $\overline{u}_2(x,y)=\max(\min(2x+y,1),2x)$ (bottom panels). The black curves in the left panels show indirect utility pairs $(v_1(p,1-p,m),v_2(p,1-p,1-m))$ for various income levels as functions of $p$. In the right panels, the red (poly) lines show the Pareto optimal allocations. The blue lines correspond to endowments that yield the same Pareto optimal allocation.}\label{fig:UPS-NC-conc}
\vspace*{-2mm}
\end{figure}
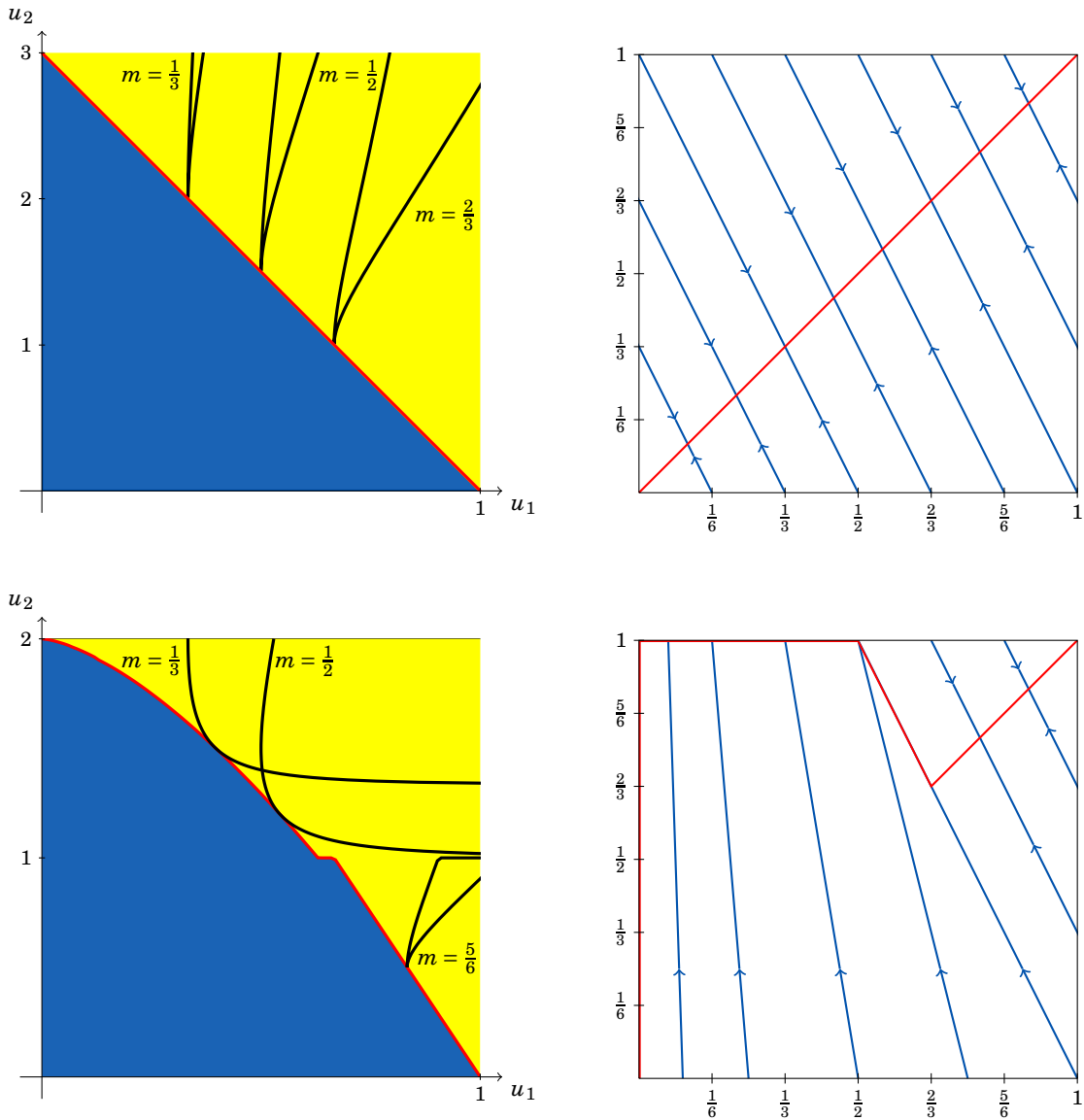

It is instructive to compare Yquilibrium with prior approaches, e.g. \cite{Starr1969} and \cite{MilgromWatt2022}, which convexify the economy by concavifying the utility functions. For $u_2(x,y)=\max(2x,y)$ the concavified utility is $\overline{u}_2(x,y)=2x+y$. The Walrasian price for the convexified economy is $p=\deel{2}{3}$. This is the price at which each of the ``V'' shaped black curves in the top-left panel of Figure \ref{fig:UPS-NC-conc} touches the frontier shared by the UPS and VPS as indicated by the red line. The top-right panel shows the Pareto-optimal allocations in the Edgeworth box. Given any income distribution $(m,1-m)$ with $0<m<1$ the Walrasian allocations of the convexified economy are $x_1=(m,m)$ and $x_2=(1-m,1-m)$, i.e. they lie on the diagonal. Obviously, the Walrasian equilibria of the convexified economy are of little use for understanding the original non-convex economy as they predict trade \textit{away} from any Pareto-optimal allocations toward inefficient allocations that are not individually rational.\footnote{For $\omega_1=(\omega,1)$ and $\omega_2=(1-\omega,0)$ the outcome is $x_1=(\deel{2}{3}\omega+\deel{1}{3},\deel{2}{3}\omega+\deel{1}{3})$ and $x_2=(\deel{2}{3}(1-\omega),\deel{2}{3}(1-\omega))$, which yields less utility for consumer 2 than the initial endowment.}

The bottom panels of Figure \ref{fig:UPS-NC-conc} shows analogous results for the quasiconcavified utilities $u_1=(x^2y)^{1/3}$ and $\overline{u}_2=\max(\min(2x+y,1),2x)$. The Pareto-optimal allocations for the resulting convex economy track those of the original non-convex economy better, but imperfectly so.\footnote{For $\overline{u}_2>1$, the quasiconcavified utility is the same as the original utility and, hence, so is the UPS. The price at which a black curve in the bottom-left panel touches the UPS' frontier is the Walrasian price for the convexified economy. When the UPS' frontier of the convexified economy matches that of the original economy, it is also the price for the non-convex economy.  This is why the blue lines in the left part of the bottom-right panel are the same as those in Figure \ref{fig:UPS-NC}. However, the blue lines in the right part correspond to a price of $p=\deel{2}{3}$, which is incorrect for the non-convex economy.}
There is still trade away from some Pareto-optimal allocations toward inefficient allocations that are not individually rational. $\hfill\blacksquare$}
\end{example}
To summarize, with few traders, convexifying the economy by (quasi) concavifying utilities can lead to counterintuitive predictions such as trading at a loss and away from Pareto-optimal to inefficient allocations. This finding contrasts with \citeauthor{MilgromWatt2022}'s (\citeyear{MilgromWatt2022}) limit result that, as the number of traders grows large, the Walrasian equilibria of the convexified economy are approximately efficient and approximately individually rational for the non-convex economy. Yquilibrium complements \citeauthor{MilgromWatt2022}'s approach in that it is explicitly non-Walrasian and produces optimal and individually rational allocations (subject to linear, anonymous prices) for any number of traders and in any (non-) convex economy.

\section{Conclusion}

This paper contributes to an emerging literature on non-convex market design. Examples include the design of the US incentive auction for radio spectrum \citep{LBMS2017} and the combinatorial exchange used for trading catch shares in New South Wales, Australia \citep{BFG2018,BFG2019}. The exchange ended two decades of political debate by providing a market-based response to a major policy problem faced by fisheries worldwide: the reallocation of catch shares in cap-and-trade programs designed to prevent overfishing. Fishers exiting the industry faced the risk of selling only part of their share portfolio, ending up with an unviable business and not enough proceeds to recoup their fixed costs. The design question was how to deal with this non-convexity using a market that employed linear and anonymous prices, which were deemed necessary for reasons of simplicity, transparency, and fairness.

The impetus for this paper was the frustration that economics' cornerstone theory, general equilibrium, could not inform this industry reform due to its axiomatic framework that assumes perfect competition and convexity. One contribution of this paper is to it shows these assumptions are not only unrealistic they are also unnecessary. For convex economies, existence of Walrasian equilibrium and the welfare theorems follow from a novel duality result for vector optimization, see Theorem \ref{dualVO} and Corollaries \ref{wt1} and \ref{wt2}.

A second contribution is that scalarizing the primal and dual vector optimization programs results in a univariate program that produces Walrasian equilibria as the maximizers, and roots, of a single function -- the economy's potential. Maximization of the potential combines earlier, and seemingly unrelated, approaches to equilibrium. \citeauthor{Negishi1960}'s (\citeyear{Negishi1960}) method is a commonly used tool in macroeconomics to obtain Pareto-optimal allocations by maximizing welfare. \cite{ausubel2006} minimizes a Lyapunov function equal to dual welfare to obtain Walrasian prices in an auction context with quasilinear preferences. Maximization of the potential, defined as the difference between welfare and its dual, produces both allocations and prices and applies to more general environments than those considered by \citeauthor{Negishi1960} and \citeauthor{ausubel2006}, see Theorem \ref{th:root} and the ensuing discussion.

The paper's main contribution is that potential maximization provides an optimization approach to non-convex markets. Formulating market outcomes via a constrained optimization problem is valuable in that it readily accommodates additional non-economic constraints, such as legal, political, or fairness constraints, which play a role in many applications. Moreover, having a solution concept that works for convex and non-convex markets alike, see Theorem \ref{potential}, is a much a needed and overdue addition to general equilibrium's incomplete toolkit.

Yquilibrium breaks with the Walrasian paradigm that puts prices front and center and instead adopts the Paretian viewpoint that gains from trade consistent with linear anonymous pricing will be seized. Prices signal what utilities can be expected from a given endowment, but they do not restrict traders behavior. In particular, Yquilibrium does not impose price taking.

Yquilibrium shows what \textit{should} happen in non-convex markets. This raises the question \textit{how} it will happen, i.e. what design ensures Yquilibrium comes about? Market design is both a science and an art and the answer to this question depends on the application. Electricity markets, spectrum auctions, and the reallocation of resource rights all require their own ``bidding language'' to reliably implement Yquilibrium.\footnote{See, e.g., \cite{BFG2018,BFG2019} for details about the bidding language used in the combinatorial exchange for catch shares in New South Wales, Australia.} 

But I would be remiss not to mention the main features of ``YCE,'' a Yquilibrium combinatorial exchange for (non-) convex exchange economies \citep{GoereeLindsayWilliams2023}.
First, YCE allows for ``all-or-nothing'' combinatorial offers that specify positive amounts of the goods that are demanded and negative amounts the goods that are offered for sale. This feature protects traders from exposure problems that can cause trade to halt in standard markets where each good is traded separately, see \cite{GoereeLindsay2020}.\footnote{Consider the ``max'' trader in the bilateral trade economy of the previous section with an initial endowment $(\deel{1}{3},1)$. The Yquilibrium outcome is that this trader ends up with the bundle $(\deel{5}{9},0)$, see Figure \ref{fig:UPS-NC}. This can be accomplished with the all-or-nothing offer $(\deel{2}{9},-1)$, which, if successful, raises the trader's utility from 1 to $\deel{10}{9}$. However, if only half of the planned trade occurs then the trader is stuck with the bundle $(\deel{4}{9},\hf)$, which yields less utility than the initial endowment. In convex markets, small trades toward the final outcome are utility improving, but this not the necessarily the case in non-convex markets. All-or-nothing combinatorial offers are needed to protect traders from losses.} Second, YCE allows for limit orders and \textit{market} orders that give traders the opportunity to accept others' orders. Third, YCE uses linear anonymous prices that follow from a simple convex minimization program. Fourth, the weights $\alpha_i$ in \eqref{Ypot} provide ``engineering'' opportunities for the designer who can set them to favor participants that provide ``liquidity'' by being willing to trade small amounts.\footnote{For instance, for the bilateral trade economy of the previous section, limit orders submitted by the Cobb-Douglas trader are typically much smaller than those submitted by the ``max'' trader.}

A first test of YCE is to compare its performance to that of the continuous double auction in non-convex exchange economies. The empirical validation of YCE, as well as its implementation in real-world applications, is left for future research.

\newpage
\addtolength{\baselineskip}{-0.23mm}
\bibliography{references}
\bibliographystyle{chicago}

\newpage

\startappendix
\addtolength{\baselineskip}{-0.9mm}

\setcounter{equation}{0}

\section{Proofs}
\label{app:proofs}

\setlength{\abovedisplayskip}{6pt}
\setlength{\belowdisplayskip}{6pt}

\noindent\textbf{Proof of Theorem 1.} The proof of \eqref{main} is based on two lemmas.

\medskip

\noindent\textbf{Lemma A1}\,\,{\em If $u\in\text{UPS}\cap\text{VPS}$ then $u$ is a maximal element of the UPS and a minimal element of the VPS.}

\medskip

\noindent\begin{proof}
If $u=(u_1,\ldots,u_N)\in\text{UPS}\cap\text{VPS}$ there exist $x\in F(w)$ and $(p,m)\in\Sigma_K(w)\times\Sigma_N$ such that $u_i=u_i(x_i)$ and $u_i=v_i(p,m_i)$ for $i\in\mathcal{N}$.
Suppose, in contradiction, that $u$ is not a minimal element of the VPS. Then there exist $(p',m')\in\Sigma_K(w)\times\Sigma_N$ such that $u_i(x_i)\geq v_i(p',m'_i)$ for $i\in\mathcal{N}$ with strict inequality for at least one $i\in\mathcal{N}$. But $v_i(p',m'_i)=u_i(x_i(p',m'_i))$ with $x_i(p',m'_i)$ consumer $i$'s optimal (Marshallian) demand that satisfies $\langle p'|x_i(p',m'_i)\rangle=m'_i$. Since indirect utility is strictly increasing in income (Lemma \ref{diewert}), we must have $\langle p'|x_i\rangle\geq m'_i$ for $i\in\mathcal{N}$ with strict inequality for at least one $i\in\mathcal{N}$, so $\sum_{i\in\mathcal{N}}\langle p'|x_i\rangle>\sum_{i\in\mathcal{N}}m'_i=1$. However, $x\in F(w)$ implies $\sum_{i\in\mathcal{N}}\langle p'|x_i\rangle\leq\langle p'|w\rangle=1$, which yields the desired contradiction.
Likewise, suppose, in contradiction, that $u$ is not a maximal element of the UPS. Then there exist $x'\in F(w)$ such that $u_i(x'_i)\geq v_i(p,m_i)$ for $i\in\mathcal{N}$ with strict inequality for at least one $i\in\mathcal{N}$. Since indirect utility is strictly increasing in income, we must have $\langle p|x'_i\rangle\geq m_i$ for $i\in\mathcal{N}$ with strict inequality for at least one $i\in\mathcal{N}$, so $\sum_{i\in\mathcal{N}}\langle p|x'_i\rangle>\sum_{i\in\mathcal{N}}m_i=1$. However, $x'\in F(w)$ implies $\sum_{i\in\mathcal{N}}\langle p|x'_i\rangle\leq\langle p|w\rangle=1$, which yields the desired contradiction.
\end{proof}

\medskip

\noindent\textbf{Lemma A2}\,\,{\em Minimal elements of the VPS belong to the UPS and maximal elements of the UPS belong to the VPS.}

\medskip

\noindent\begin{proof}
Minimal elements of the VPS can be obtained from the program
\begin{displaymath}
  \min_{{\rule{0pt}{6pt}(p,m)\,\in\,\Sigma_K(w)\times\Sigma_N}\atop{\rule{0pt}{6pt}v_i(p,m_i)\,\leq\,\overline{v}_i\,\forall\,i\,<\,N}}v_N(p,m_N)
\end{displaymath}
The first-order conditions with respect to prices and incomes are
\begin{eqnarray*}
  \nabla_p v_N+\sum_{i\,<\,N}\lambda_i\nabla_p v_i &=& \nu\;\;\;\;\;\;\forall\,k\,\in\,\mathcal{K}\\[1mm]
  -\lambda_i\frac{\partial v_i}{\partial m_i} &=& \nu\;\;\;\;\;\;\forall\,i\,<\,N\\[1mm]
  -\frac{\partial v_N}{\partial m_N} &=& \nu
\end{eqnarray*}
where $\lambda_i$ is the multiplier for $v_i(p,m_i)\leq\overline{v}_i$ and $\nu$ is the multiplier for the homogeneity-zero constraint $\sum_{i\in\mathcal{N}}m_i=\sum_{k\in\mathcal{K}}p_k$. (Gradients are replaced with (Clarke) subdifferentials if an indirect utility is not differentiable.) Combined the first-order conditions imply
\begin{displaymath}
  w+\sum_{i\,\in\,\mathcal{N}}\,\Bigl(\frac{\partial v_i}{\partial m_i}\Bigr)^{-1}\nabla_p v_i\,=\,0
\end{displaymath}
Now $v_i(p,m_i)=u_i(x_i(p,m_i))$ where consumer $i$'s optimal demand $x_i(p,m_i)$ satisfies Roy's identity: $x_{i}(p,m_i)=-\nabla_p v_i/\partial_{m_i}v_i$. The first-order conditions thus imply
\begin{displaymath}
  \sum_{i\,\in\,\mathcal{N}}x_{i}(p,m_i)\,=\,w
\end{displaymath}
In other words, if $v=(v_1(p,m_1),\ldots,v_N(p,m_N))$ is a minimal element of the VPS then $v=(u_1(x_1(p,m_1)),\ldots,u_N(x_N(p,m_N)))$ where the optimal demands $x_{i}(p,m_i)$ for $i\in\mathcal{N}$ are feasible. Hence, $v$ belongs to the UPS.

To show that any maximal element $u=(u_1(x_1),\ldots,u_N(x_N))$ of the UPS belong to the VPS recall that, for $x=(x_1,\ldots,x_N)\in F(w)$,
\begin{equation}\label{u}
  u_i(x_i)\,=\,\min_{\langle p_i|x_i\rangle\,=\,m_i}\,v_i(p_i,m_i)
\end{equation}
If $u$ is maximal then $\sum_{i\in\mathcal{N}}x_i=w$ and $\sum_{i\in\mathcal{N}}\langle p|x_i\rangle=\sum_{i\in\mathcal{N}}m_i=1$. What I need to show is that if $u$ is maximal then the price $p_i$ that minimizes consumer $i$'s indirect utility is the same for all $i\in\mathcal{N}$.

A standard envelope argument applied to \eqref{u} shows that $\partial u_i/\partial {x_{ik}}=(\partial v_i/\partial m_i)p_{ik}$ if $x_{ik}>0$ and $\partial u_i/\partial {x_{ik}}\leq(\partial v_i/\partial m_i)p_{ik}$ if $x_{ik}=0$. Let $\text{MRS}^i_{kl}=\partial_{x_{ik}}u_i/\partial_{x_{il}}u_i$ denote consumer $i$'s marginal rate of substitution. There exists a common price vector, $p$, that minimizes every consumer's indirect utility if, for $i\in\mathcal{N}$ and $k,l\in\mathcal{K}$,
\begin{equation}\label{FOC}
  \left\{\begin{array}{lll}
  \text{MRS}^i_{kl}\,=\,\frac{p_k}{p_l} & \text{if} & x_{ik}\,>\,0,\,x_{il}\,>\,0 \\[1mm]
  \text{MRS}^i_{kl}\,\leq\,\frac{p_k}{p_l} & \text{if} & x_{ik}\,=\,0,\,x_{il}\,>\,0 \\[1mm]
  \text{MRS}^i_{kl}\,\geq\,\frac{p_k}{p_l} & \text{if} & x_{ik}\,>\,0,\,x_{il}\,=\,0
  \end{array}\right.
\end{equation}
Consider the program to obtain maximal elements of the UPS:
\begin{displaymath}
  \max_{{\rule{0pt}{6pt}x\,\in\,F(w)}\atop{\rule{0pt}{6pt}u_i(x_i)\,\geq\,\overline{u}_i\,\forall\,i\,<\,N}}\,u_N(x_N)
\end{displaymath}
The first-order conditions are
\begin{eqnarray*}
  \lambda_i\frac{\partial u_i}{\partial x_{ik}} &=& \mu_k-\nu_{ik}\;\;\;\;\;\;\forall\,i\,<\,N \\[1mm]
  \frac{\partial u_N}{\partial x_{Nk}} &=& \mu_k-\nu_{Nk}
\end{eqnarray*}
where $\lambda_i$ is the multiplier for $u_i(x_i)\geq\overline{u}_i$, $\mu_k$ is the multiplier for the feasibility constraint $\sum_{i\in\mathcal{N}}x_{ik}\leq w_k$, and $\nu_{ik}$ is the multiplier for the non-negativity constraint $x_{ik}\geq 0$. I next show the first-order conditions imply \eqref{FOC}. If $x_{ik}>0$ and $x_{il}>0$ then $\nu_{ik}=\nu_{il}=0$ and $\text{MRS}^i_{kl}=\mu_k/\mu_l$. If $x_{ik}=0$ and $x_{il}>0$ then $\nu_{ik}\geq 0$ and $\nu_{il}=0$ so $\text{MRS}^i_{kl}\leq\mu_k/\mu_l$. Finally, if $x_{ik}>0$ and $x_{il}=0$ then $\nu_{ik}=0$ and $\nu_{il}\geq 0$ so $\text{MRS}^i_{kl}\geq\mu_k/\mu_l$. To summarize, the conditions in \eqref{FOC} are satisfied for $p=\mu$.
\end{proof}

\noindent To show that the solution to \eqref{main} is unique for a given $m\in\Sigma_N$, suppose, in contradiction, that $(p,m)$ and $(p',m)$ are two solutions to \eqref{main}. This means that both are minimal elements of the VPS. Strict quasiconcavity of the indirect utilities implies that, for $i\in\mathcal{N}$, $v_i(\hf(p+p'),m)<v_i(p,m)$ or $v_i(\hf(p+p'),m)<v_i(p',m)$, contradicting minimality of either $(p,m)$ or $(p',m)$.$\hfill\blacksquare$

\medskip

\noindent\textbf{Proof of Theorem 2.} For allocations that satisfy consumers' budget constraints, $u_i(x_i)\leq v_i(p,\langle p|\omega_i\rangle)=
\max_{\langle p|x'_i\rangle=\langle p|\omega_i\rangle}u_i(x'_i)$ for $i\in\mathcal{N}$, so the program's value is non-positive. If $(x,p)$ is a Walrasian equilibrium then allocations are optimal at price $p$, i.e. $u_i(x_i)=\max_{\langle p|x'_i\rangle=\langle p|\omega_i\rangle}u_i(x'_i)$, so $(x,p)$ is a root, whence maximizer. Conversely, if $(x,p)$ is a root of the objective then each term in the objective's sum is zero (as they are all non-positive and weights are positive). But $u_i(x_i)=v_i(p,\langle p|\omega_i\rangle)$ for $i\in\mathcal{N}$ means everyone is maximizing at price $p$ and $x\in F(w)$ means the optimal demands are feasible. Hence, $(x,p)$ is a Walrasian equilibrium. $\hfill\blacksquare$

\medskip

\noindent\textbf{Proof of Theorem 3.} Consider a feasible allocation $x\in F(w)$ that satisfies budget constraints: $\langle p|x_i\rangle=\langle p|\omega_i\rangle$ for $i\in\mathcal{N}$. Since $u_i(x_i)\leq v_i(p,\langle p|\omega_i\rangle)=\max_{\langle p|x'_i\rangle=\langle p|\omega_i\rangle}u_i(x'_i)$ for $i\in\mathcal{N}$, the value of the program in \eqref{pot} is non-positive. Suppose the economy has a Walrasian equilibrium $(x,p)$ then allocations are optimal at price $p$, i.e. $u_i(x_i)=\max_{\langle p|x'_i\rangle=\langle p|\omega_i\rangle}u_i(x'_i)=v_i(p,m_i)$ for $i\in\mathcal{N}$, so $(x,p)$ is a root, whence maximizer, of the potential. Hence, any Walrasian equilibrium is a Yquilibrium. I next show there can be no Yquilibrium $(x,p)$ different from any of the Walrasian equilibria. The latter produce a zero value for the program in \eqref{pot} and, hence, so must the Yquilibrium. But, if $(x,p)$ is a root of the objective then each term in the objective's sum is zero (as they are all non-positive and weights are positive). And $u_i(x_i)=v_i(p,\langle p|\omega_i\rangle)$ for $i\in\mathcal{N}$ means everyone is maximizing at price $p$ and $x\in F(w)$ means the optimal demands are feasible. Hence, $(x,p)$ is a Walrasian equilibrium.

When Walrasian equilibrium does not exist the properties of Yquilibria listed in Theorem 3 follow by construction of the maximization program in \eqref{pot}.$\hfill\blacksquare$

\vspace*{-3mm}
\section{The Dual Negishi Program}
\label{app:dualNegishi}

The dual program entails minimizing dual welfare $V_\alpha(p,m)=\sum_{i\in\mathcal{N}}\alpha_iv_i(p,m_i)$ with respect to prices
\begin{equation}\label{W}
  W^*_\alpha(w,m)\,=\,\min_{\rule{0pt}{7pt}\langle p|w\rangle\,=\,\sum_im_i}\,V_\alpha(p,m)
\end{equation}
The first step is to determine the price as a function of the weights. Strict quasiconvexity of the indirect utilities ensures the price is unique. Given the price, consumers' optimal demands follow from Roy's identity. The next step is to solve for weights that yield a price vector such that optimal demands are feasible. The correct weights are the same as in \citeauthor{Negishi1960}'s (\citeyear{Negishi1960}) program, i.e. $\alpha_i=1/(\partial v_i/\partial m_i)$ for $i\in\mathcal{N}$.
\begin{theorem}\label{gradient}
Walrasian equilibrium prices solve \eqref{W} provided welfare weights are equal to the inverse marginal utilities of income. Walrasian prices satisfy
\begin{equation}\label{Grad}
  p\,=\,\nabla_w W^*_\alpha(w,m)
\end{equation}
which reflects that they are the dual variables of the resource constraints.
\end{theorem}

\medskip

\noindent\textbf{Proof of Theorem 4.} The Lagrangian for the minimization program is
\begin{displaymath}
  \sum_{i\,\in\,\mathcal{N}}\alpha_iv_i(p,m_i)+\lambda\bigl(\langle p|w\rangle-\sum\nolimits_{i\,\in\,\mathcal{N}}m_i\bigr)
\end{displaymath}
The first term is homogeneous of degree zero in income and prices and the second term is homogeneous of degree one. I use this to normalize $\lambda=1$.  The first-order condition is
\begin{displaymath}
  w+\sum_{i\,\in\,\mathcal{N}}\alpha\nabla_pv_i(p,m_i)\,=\,0
\end{displaymath}
Using $\alpha_i=1/(\partial v_i/\partial m_i)$ and Roy's identity this can be rewritten as
\begin{displaymath}
  \sum_{i\,\in\,\mathcal{N}}x_i(p,m_i)\,=\,w
\end{displaymath}
Hence, the first-order condition for minimization implies the price is such that consumers' optimal bundles are feasible, i.e. $p$ is a Walrasian equilibrium price. Finally, a simple envelope argument establishes $p=\nabla_w W^*_\alpha(w,m)$. $\hfill\blacksquare$

\begin{example}{\em
Consider an economy with Cobb-Douglas utilities $u_1=x\sqrt{y}$ and $u_2=\sqrt{x}y$ and incomes $(m_1,m_2)$. The blue area in the left panel of Figure \ref{fig:UPS} shows the UPS, which is not convex because the Cobb-Douglas utilities exhibit increasing returns to scale and are not concave. As a result, the usual method of identifying Pareto optimal allocations by maximizing a weighted some of utilities over the UPS does not apply.

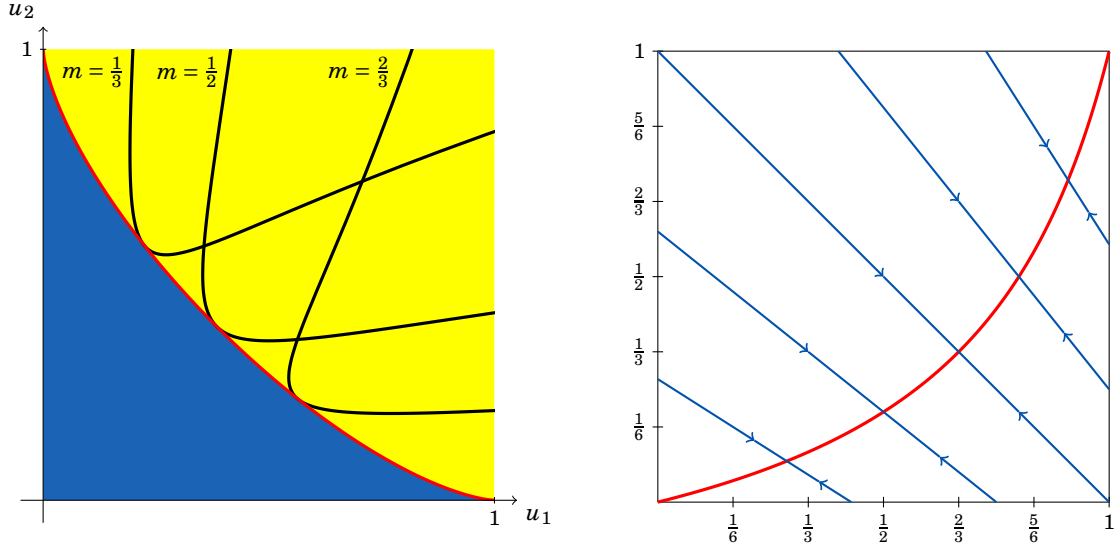
\begin{figure}[t]
\begin{center}
\begin{tikzpicture}[scale=6]

\begin{scope}

\clip (0,0) rectangle (1,1);

\draw [name path=A, very thin, ukrainianBlue, domain=0:1, samples=101, /pgf/fpu,/pgf/fpu/output format=fixed] plot ({2*\x/(1+\x)*(\x/(2-\x))^.5}, {(1-2*\x/(1+\x))^.5*(1-(\x/(2-\x)))});
\draw [name path=B] (0,0);
\draw [name path=C] (1,1);
\tikzfillbetween[of=A and B]{ukrainianBlue, opacity=0.9};
\tikzfillbetween[of=A and C]{ukrainianYellow, opacity=1};

\draw [very thick, black, domain=0.01:0.99, samples=101, /pgf/fpu,/pgf/fpu/output format=fixed] plot ({2/27*1/\x*1/(1-\x)^.5}, {4/27*1/(1-\x)*(2/\x)^.5});
\draw [very thick, black, domain=0.01:0.99, samples=101, /pgf/fpu,/pgf/fpu/output format=fixed] plot ({1/54^.5*1/\x*1/(1-\x)^.5}, {1/54^.5*1/(1-\x)*(1/\x)^.5});
\draw [very thick, black, domain=0.01:0.99, samples=101, /pgf/fpu,/pgf/fpu/output format=fixed] plot ({4/27*2^.5/(1-\x)^.5*(1/\x)}, {2/27*1/\x^.5*1/(1-\x)});

\node[scale=0.8,black] at (0.11,0.95) {$m=\deel{1}{3}$};
\node[scale=0.8,black] at (0.32,0.95) {$m=\deel{1}{2}$};
\node[scale=0.8,black] at (0.7,0.95) {$m=\deel{2}{3}$};

\draw [very thick, red, domain=0:1, samples=101, /pgf/fpu,/pgf/fpu/output format=fixed] plot ({2*\x/(1+\x)*(\x/(2-\x))^.5}, {(1-2*\x/(1+\x))^.5*(1-(\x/(2-\x)))});

\end{scope}

\draw[->] (-0.05,0) -- (1.05,0) node[scale=0.8,below right] {$u_1$};
\draw[->] (0,-0.05) -- (0,1.05) node[scale=0.8,above left] {$u_2$};
\foreach \x/\xtext in {1/1} \draw[shift={(\x,0)}] (0pt,.2pt) -- (0pt,-.2pt) node[scale=0.7,below] {$\xtext$};
\foreach \y/\ytext in {1/1} \draw[shift={(0,\y)}] (.2pt,0pt) -- (-.2pt,0pt) node[scale=0.7,left] {$\ytext$};

\node at (1.84,0.46) {
\begin{tikzpicture}[scale=1]
\draw (0,0) -- (6,0) -- (6,6) -- (0,6) -- (0,0);
\foreach \x/\xtext in {1/\frac{1}{6}, 2/\deel{1}{3}, 3/\hf, 4/\deel{2}{3}, 5/\deel{5}{6}, 6/1} \draw[shift={(\x,0)}] (0pt,2pt) -- (0pt,-2pt) node[scale=0.7,below] {$\xtext$};
\foreach \y/\ytext in {1/\frac{1}{6}, 2/\deel{1}{3}, 3/\hf, 4/\deel{2}{3}, 5/\deel{5}{6}, 6/1} \draw[shift={(0,\y)}] (2pt,0pt) -- (-2pt,0pt) node[scale=0.7,left] {$\ytext$};

\draw [very thick, red, domain=0:1, samples=101] plot ({12*\x/(1+\x)}, {6*\x/(2-\x)});

\draw [thick, ukrainianBlue, domain=0:1.5/7, ->] plot ({6*\x}, {6*(3-7*\x)/11});
\draw [thick, ukrainianBlue, domain=1.5/7:2.5/7] plot ({6*\x}, {6*(3-7*\x)/11});
\draw [thick, ukrainianBlue, domain=2.5/7:3/7,<-] plot ({6*\x}, {6*(3-7*\x)/11});

\draw [thick, ukrainianBlue, domain=0:1/3,->] plot ({6*\x}, {6*(3-4*\x)/5});
\draw [thick, ukrainianBlue, domain=1/3:5/8] plot ({6*\x}, {6*(3-4*\x)/5});
\draw [thick, ukrainianBlue, domain=5/8:3/4,<-] plot ({6*\x}, {6*(3-4*\x)/5});

\draw [thick, ukrainianBlue, domain=0:1/2,->] plot ({6*\x}, {6*(3-3*\x)/3});
\draw [thick, ukrainianBlue, domain=1/2:4/5] plot ({6*\x}, {6*(3-3*\x)/3});
\draw [thick, ukrainianBlue, domain=4/5:1,<-] plot ({6*\x}, {6*(3-3*\x)/3});

\draw [thick, ukrainianBlue, domain=2/5:2/3,->] plot ({6*\x}, {6*(6-5*\x)/4});
\draw [thick, ukrainianBlue, domain=2/3:9/10] plot ({6*\x}, {6*(6-5*\x)/4});
\draw [thick, ukrainianBlue, domain=9/10:1,<-] plot ({6*\x}, {6*(6-5*\x)/4});

\draw [thick, ukrainianBlue, domain=8/11:9.5/11,->] plot ({6*\x}, {6*(15-11*\x)/7});
\draw [thick, ukrainianBlue, domain=9.5/11:10.5/11] plot ({6*\x}, {6*(15-11*\x)/7});
\draw [thick, ukrainianBlue, domain=10.5/11:1,<-] plot ({6*\x}, {6*(15-11*\x)/7});

\end{tikzpicture}};

\end{tikzpicture}
\end{center}
\vspace*{-6mm}

\caption{In the left panel, the blue area corresponds to the UPS and the yellow area to the VPS for $u_1(x,y)=xy^{1/2}$ and $u_2(x,y)=x^{1/2}y$. The sets overlap only along their frontiers indicated by the red curve. The black curves show the indirect utility pairs $(v_1(p,1-p,m),v_2(p,1-p,1-m))$ for various income levels as functions of $p$. In the right panel, the red curve shows Pareto optimal allocations in the Edgeworth box and the blue lines show endowments resulting in the same Pareto optimal allocation.}\label{fig:UPS}
\vspace*{-2mm}
\end{figure}

The first-order condition for dual welfare minimization is $-\nabla_p V_\alpha(p,m)=(1,1)$ where the right side arises from the constraint $p+q=m_1+m_2$. (Dual welfare can be made arbitrarily small if incomes are fixed and prices are unrestricted.) Now
\begin{displaymath}
  -\nabla_p V_\alpha(p,m)\,=\,\Bigl(\,\frac{2\alpha_1m_1\sqrt{m_1q}+\alpha_2m_2\sqrt{m_2p}}{3\sqrt{3}p^2q},\frac{\alpha_1m_1\sqrt{m_1q}+2\alpha_2m_2\sqrt{m_2p}}{3\sqrt{3}pq^2}\,\Bigr)
\end{displaymath}
and substituting $\alpha_1=1/(\partial v_1/\partial m_1)=p\sqrt{3q/m_1}$ and $\alpha_2=1/(\partial v_2/\partial m_2)=q\sqrt{3p/m_2}$ the right side simplifies to
\begin{displaymath}
  -\nabla_p V_\alpha(p,m)\,=\,\Bigl(\,\frac{2m_1+m_2}{3p},\frac{m_1+2m_2}{3q}\,\Bigr)
\end{displaymath}
and equating it to $(1,1)$ yields the Walrasian equilibrium prices.
$\hfill\blacksquare$}
\end{example}
Minimizing dual welfare also applies when the economy is parameterized by endowments rather than incomes. Define $V_\alpha(p,\omega)=\sum_{i\in\mathcal{N}}\alpha_iv_i(p,\langle p|\omega_i\rangle)$ where $\omega$ is the concatenation of consumers' endowment vectors. Consider the program
\begin{equation}\label{W2}
  W^*_\alpha(\omega)\,=\,\min_{\rule{0pt}{7pt}p\,\in\,\Sigma_K(w)}\,V_\alpha(p,\omega)
\end{equation}
Incomes are no longer fixed and the objective in \eqref{W2} is homogeneous of degree zero in prices. The normalization $\langle p|w\rangle=1$ no longer causes a binding constraint, and the first-order condition is therefore $\nabla_p V_\alpha(p,\omega)=0$.
\addtocounter{example}{-1}
\begin{example}[continued]{\em
Suppose consumers' endowments are $\omega_1=(\omega_{11},\omega_{12})$ and $\omega_2=(1-\omega_{11},1-\omega_{12})$. Evaluated at weights $\alpha_i=1/(\partial v_i/\partial m_i)$ for $i=1,2$ the gradient is
\begin{displaymath}
  -\nabla_p V_\alpha(p,\omega)\,=\,\Bigl(\frac{1+\omega_{12}-p(3-\omega_{11}+\omega_{12})}{3p},\frac{1+\omega_{12}-p(3-\omega_{11}+\omega_{12})}{3(1-p)}\Bigr)
\end{displaymath}
which vanishes when $p=(1+\omega_{12})/(3-\omega_{11}+\omega_{12})$.
$\hfill\blacksquare$}
\end{example}
Finally, the dual welfare minimization program is more efficient than \citeauthor{Negishi1960}'s (\citeyear{Negishi1960}) welfare maximization program. It involves $K$ parameters, i.e. the prices $p_k$ for $k\in\mathcal{K}$, instead of $NK$ parameters, i.e. the individual allocations $x_{ik}$ for $i\in\mathcal{N}$, $k\in\mathcal{K}$.

\end{document}